\documentclass[useAMS,usenatbib,twocolumn]{mn2e}
\usepackage{xspace}
\usepackage{hyperref}
\usepackage{aas_macros}
\usepackage{url}
\usepackage{amsmath}
\usepackage{color}
\usepackage{amssymb}
\usepackage[capitalise]{cleveref}
\usepackage{graphicx}
\usepackage{subfig}
\usepackage{enumitem}

\usepackage{etoolbox}
\makeatletter
\patchcmd\@combinedblfloats{\box\@outputbox}{\unvbox\@outputbox}{}{%
   \errmessage{\noexpand\@combinedblfloats could not be patched}%
}%
 \makeatother

\newcommand{\E}{\times10}
\newcommand{\jbca}{{Jodrell Bank Centre for Astrophysics, School of Physics \& Astronomy, The University of Manchester, Manchester M13 9PL, UK}}
\newcommand{\icg}{{Institute of Cosmology \& Gravitation, University of Portsmouth,
Dennis Sciama Building, Burnaby Road, Portsmouth PO1 3FX, UK}}

\newcommand{\qmu}{{School of Physics and Astronomy, Queen Mary University of London, Mile End Road, London E1 4NS, UK}}

\newcommand{\euclid}{\textit{Euclid}\xspace}

\newcommand{\sax}{S$^3$-SAX\xspace}
\newcommand{\atoa}{A$^2$A\xspace}
\newcommand{\hi}{H\,\textsc{i}\xspace}
\newcommand{\tocm}{\ensuremath{21\,\mathrm{cm}}\xspace}
\title[HI Intensity Mapping for Clustering-Based Redshift Estimation]{HI Intensity Mapping for Clustering-Based Redshift Estimation}

\author[S.Cunnington et al.]{Steven Cunnington$^{1}$\thanks{E-mail: steve.cunnington@port.ac.uk}, Ian Harrison$^{2}$, Alkistis Pourtsidou$^{3,1}$, David Bacon$^{1}$\\
$^{1}$\icg\\
$^{2}$\jbca\\
$^{3}$\qmu\\
}

\begin{document}
%\date{Accepted 1988 December 15. Received 1988 December 14; in original form 1988 October 11}

\pagerange{\pageref{firstpage}--\pageref{lastpage}} \pubyear{2018}

\maketitle

\label{firstpage}

\begin{abstract}
 Precision cosmology requires accurate galaxy redshifts, but next generation optical surveys will observe unprecedented numbers of resolved galaxies, placing strain on the amount of spectroscopic follow-up required. We show how useful information can be gained on the redshift distributions of optical galaxy samples from spatial cross-correlations with intensity maps of unresolved \hi (21cm) spectral line emission. We construct a redshift distribution estimator, which we test using simulations. We utilise the S$^3$-SAX catalogue which includes \hi emission information for each galaxy, which we use to construct \hi intensity maps. We also make use of simulated LSST and \euclid-like photometry enabling us to apply the \hi clustering calibration to realistic simulated photometric redshifts. While taking into account important limitations to \hi intensity mapping such as lost $k$-modes from foreground cleaning and poor angular resolution due to large receiver beams, we show that excellent constraints on redshift distributions can be provided for an optical photometric sample.
\end{abstract}


%----------------------------------------------------------------------------------------
\section{Introduction}
%----------------------------------------------------------------------------------------
\label{sec:introduction}

According to the standard cosmological model, dark energy
is responsible for the current acceleration of the Universe's expansion \citep{ReissAcceleratingUniverse, PerlmutterAcceleratingUniverse}. The next step towards constraining our cosmological model relies on precise measurements of the 3-dimensional large-scale structure. The majority of this structure is in the form of underlying dark matter which does not interact with light and is therefore invisible to our telescopes. However, making the well reasoned assumption that light emitting galaxies act as a biased tracer of this underlying dark matter distribution, we can use large optical surveys to construct catalogues of galaxies. We then process and analyse these catalogues to construct a 3-dimensional map of the Universe.
This relies heavily on having a good method for measuring the radial distance out to all these galaxies, i.e. having a good estimate of the galaxy redshifts.

There exist two approaches to measuring redshifts in optical catalogues, spectroscopy and photometry. Spectroscopy is the more accurate of the two but is time consuming since it relies on gathering a large number of photons for any one galaxy. An estimation of redshift is then obtained by observation of known emission or absorption lines in the spectral energy distribution (SED). With the rapidly increasing orders of magnitude of galaxy numbers detected by forthcoming surveys such as the Large Synoptic Survey Telescope\footnote{www.lsst.org} (LSST) and \euclid\footnote{www.euclid-ec.org}-like surveys, a time-expensive method such as spectroscopy is unlikely to be a viable method for measuring the redshift for these large populations.

Often surveys need to settle for the photometry approach \citep[e.g.][]{PhotozIntro}, which is faster but not as accurate as spectroscopy. This relies on obtaining the SED from broad-band photometry i.e. measuring the amount of flux collected in each of the telescope's broad colour filters, and relies on strong galaxy spectral features such as the 4000\AA\ break being detectable. Obtaining photometric redshifts can therefore be thought of as spectroscopy with extremely low resolution; for example the LSST plans to operate with six colour filters, \textit{ugrizy} \citep{LSSTwhitepaper}. Photometric redshift methods can generally be categorised into either template fitting methods, where various spectrum templates are fitted to find a close match, or opting for machine learning methods where a training set is used to derive a relation between redshifts and colour magnitudes \citep{MachineLearningANNz2}. Opting for a photometric approach means a far greater number of galaxies can have estimated redshifts, but more detailed consideration must be taken of the redshift error associated with this technique.

A method to calibrate photometric redshifts, without the need for verification from time-expensive spectroscopic follow-up, is to use clustering-based redshift estimation. The general idea is to use a pre-existing `reference' sample for which some precise redshift information has already been gained, and which spatially overlaps with the photometric sample which can be treated as having unknown redshift. Then by utilising the spatial clustering of galaxies within the overlapping samples through cross-correlations, we can constrain the `unknown' (photometric) redshift distribution. In other words, where there is strong angular clustering between the unknown sample and a slice of the known sample at a particular redshift, one can infer that the unknown sample is well represented in that particular redshift bin. From this principle we can build an estimated redshift distribution for the unknown sample, giving much more constrained redshift information for the particular population of galaxies. 

There is now a significant amount of literature on clustering-based redshift estimation, with \citet{Newman08Clustering-z} being one of the first to demonstrate the method on simulations. The method has since been refined with simulations by \citet{Matthews&NewmanClustering-z, BenjaminClustering-z, SchmidtClustering-z, vanDaalenClusteringz}, and others have more recently applied the approach to real data \citep{MenardClustering-z, RahmanClusteringz}. Most recently the Dark Energy Survey have applied the clustering redshifts method to their Year 1 Data in \citet{DES1clusteringz} and \citet{DES2clusteringz}.

The appeal of this idea is that when LSST and \euclid-like surveys deliver unprecedented galaxy catalogue sizes but lack well-constrained redshift information, we do not need to rely on time-consuming spectroscopic follow-up on every galaxy, or representative sub-samples which are not biased with respect to the full survey. Instead we can utilise a pre-existing, spatially overlapping, catalogue for which there is precise redshift information and use this as the `reference' sample in a clustering-based redshift estimation.

However, there is no reason why the reference sample needs to be a sample of resolved galaxies. The idea should work just as well if one cross-correlates with any tracer of large scale structure. The idea that we will investigate in this paper is the use of unresolved maps of neutral hydrogen (\hi), obtained through the technique of intensity mapping \citep{Battye2004, 21cmIntro, LineIMStatusReport}. Intensity maps and photometric galaxy surveys are highly complementary to one another, with photometric surveys having high spatial but low spectral resolution, and intensity maps high spectral but low spatial resolution.

Intensity mapping works by isolating a particularly clear spectral feature e.g. the \hi line (or \tocm line) in radio observations, and then associating the observed intensity at this frequency with the appropriate redshift based on this feature. Intensity mapping is not concerned with resolving individual galaxies (i.e. it will typically operate with poor angular resolution), but has excellent redshift information from the combined emission of numerous sources, including faint ones that could well go undetected in conventional optical surveys.

The \hi \tocm line is due to the energy emitted when a neutral hydrogen atom undergoes the hyperfine spin flip. Although this is a rare event with a single atom taking on average $10^7$ years to undergo this process, we are helped by the fact that there is still a huge abundance of neutral hydrogen in our late-time Universe, making this faint signal detectable and a tracer of the underlying large scale structure \citep{PenIMfirstdetection, ParkesIMxOptDetection}. The radiation emitted from this process carries a rest wavelength of \tocm (1420 MHz) and hence its redshifted signal falls within the bounds of radio astronomy. 

Intensity mapping can theoretically be carried out on any spectral line, however \hi intensity mapping is the common choice, especially for cosmologists. This is firstly because there is no other dominant spectral line close to its emitted frequency, therefore avoiding any line confusion from other spectral features. Secondly,  even though in the epoch of reionization (approximately $6\leq z \leq 15$) it is thought that the power spectrum measured from \hi intensity mapping will be largely shaped by the pattern of ionized regions, in the post-reionization era i.e. once reionization is complete ($z<6$), some \hi will still remain in collapsed objects and the \hi power spectrum will therefore be a measure of the underlying matter power spectrum \citep{PostReionizationWythe}.

Intensity mapping instruments can generally be drawn into two categories, either operating as a single-dish receiver or as an interferometer \citep{,BullLateTime21cm}. For single-dish experiments such as BINGO \citep{BattyeBINGOSingleDish,BINGOupdate} or FAST \citep{FAST} the signal is incident on one receiver and auto-correlated. For interferometers such as CHIME \citep{CHIME} or HIRAX \citep{HIRAX} the signal is incident on a set of receivers and cross-correlated. Arguably one of the most exciting upcoming radio surveys is the Square Kilometre Array (SKA) \citep{SKAHICosmologyPoS}. The MeerKLASS survey using the SKA's pathfinder MeerKAT \citep{MeerKLASSMeerKAT, Pourtsidou:2017era} is expected to be capable of \hi intensity mapping in both interferometric and single-dish modes.

For detections to be possible by these instruments, careful consideration needs to be given to radio foregrounds; for example, synchrotron radiation coming from our own Milky Way can dominate over the faint \hi signal by several orders of magnitude. The highest redshift detection of \hi emission in a targeted single object is currently at $z=0.376$ \citep{HIhighzdetection}. At higher redshifts the emission is too weak for detection with current instruments; however, detections are possible through 21 cm absorption signals \citep{HighzHIabsorption}. In addition, measurements of the \hi cosmological mass density $\Omega_\text{HI}$ have been made at high redshifts \citep{HighzHIMassDensity} but there it is inferred from damped Ly-$\alpha$ systems that trace the bulk of neutral gas in the universe. Detections of \hi using intensity mapping have been made using cross-correlations with optical galaxy surveys; the first detection of cosmic structure using \hi intensity maps was reported in \citet{PenIMfirstdetection,Chang:2010jp}. The only claimed measurements of a power spectrum obtained using \hi intensity mapping are from the Green Bank Telescope ($0.6<z<1$) \citep{GBTHIdetection1, GBTHIdetection2} and the Parkes Observatory ($0.057<z<0.098$) \citep{ParkesIMxOptDetection}, which again relied on cross-correlations with optical surveys to boost their signal. These cross-correlation detections have a particular importance in that they prove that the \hi emission signal correlates with optical galaxies which are known tracers of the underlying matter distribution. This therefore justifies using \hi intensity mapping as a technique for probing large-scale structure and, consequently, for clustering-based redshift estimation.

It is apparent that cross-correlations can be beneficial for both optical surveys and radio \hi intensity mapping experiments. Radio can help calibrate  photometric redshifts, and optical galaxy surveys can help radio \hi intensity mapping surveys by mitigating systematic effects and residual foreground contamination \citep{AlkistisIMoptcross, AlkistisIMoptCMBcross}.

In this paper we aim to extend previous work by \citet{AlonsoIMClusteringz} and investigate the use of \hi intensity maps for clustering-based redshift estimation. We will be taking a simulation based approach and attempting to recover the redshift distribution for an optical galaxy catalogue that we will be treating as our `unknown' redshift sample. This will be done through cross-correlations with \hi intensity maps (our `reference' sample) which we simulate from the same catalogue so that they share a clustering signal. We can then compare our estimated redshift distribution with the true distribution of that catalogue.

Our paper is laid out as follows: in Section \ref{SimDetails} we outline the simulation recipe used for producing our \hi intensity maps and optical galaxy count maps. This also includes how we make use of, and extend upon, catalogues with simulated optical photometry for LSST and \euclid-like experiments, which we later use in comparisons with our estimated redshift information. Section \ref{EstimatorFormalism} goes through the derivation of our estimator, which we will be using for predicting the redshift distribution of the optical sample. In Section \ref{Results} we present our findings for the clustering approach to redshift estimation, and discuss its impact and limitations. We conclude and summarise in Section \ref{Conclusion}.

%----------------------------------------------------------------------------------------
\section{Simulations}\label{SimDetails}
%----------------------------------------------------------------------------------------

For this work we principally make use of the S$^3$-SAX simulation \citep{SAXObreschkow} for investigating the limitations of using \hi intensity maps for clustering-based redshift estimation. However, we also make use of other simulations depending on the specific requirements of our tests. When seeking to demonstrate the calibration capability on photometric redshifts we require a catalogue which has robustly simulated photometry (discussed in Section \ref{PhotozSec}). When seeking to test our \hi intensity maps at low resolutions we require a simulation covering a much larger sky area (discussed in Section \ref{LargeSky}).

\begin{figure}
	\includegraphics[width=\columnwidth]{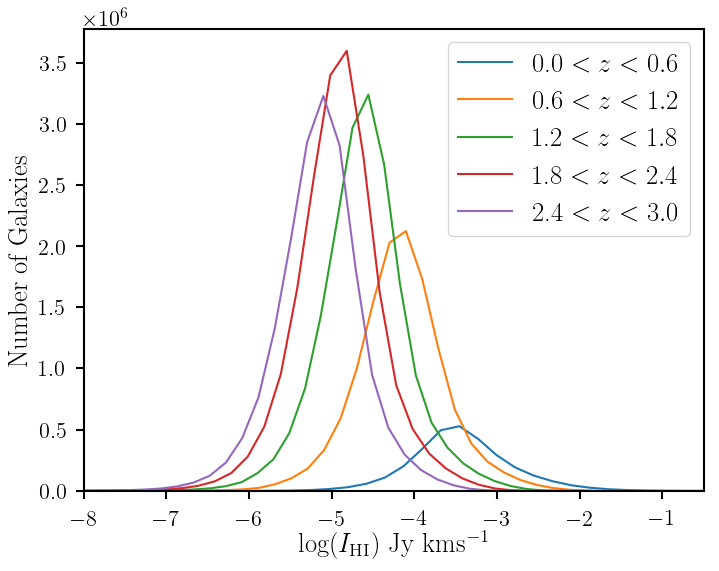}
    \caption{\hi flux histograms for galaxies in the S$^3$-SAX catalogue for different redshift bins. This shows the range of fluxes which will contribute to our \hi intensity maps.}
    \label{SAXfluxHistogram}
\end{figure}

We begin by discussing the \sax catalogue which is used for the majority of this work since it contains simulated \hi information for all its galaxies.
This is a semi-analytic simulation of a sky field with apparent \hi emission properties for approximately $2.8 \times 10^8$ galaxies in a virtual observing cone whose properties have been derived from the Millennium dark matter simulations \citep{MillenniumSimulation}. The catalogue we extract from S$^3$-SAX contain galaxies spanning 36 ${\rm deg}^2$ and extends up to a redshift of $z=3$, which approximately covers the redshift range of forthcoming stage-IV photometric telescopes which could benefit from our type of clustering-based redshift estimation. From this catalogue we use the columns for right ascension, declination, apparent redshift (which includes peculiar velocities), and \hi-mass.

By using a galaxy catalogue from a simulation like this, we can construct realistic \hi intensity maps from the integrated effect of apparent properties of each contributing galaxy. Furthermore, since the S$^3$-SAX catalogue already considers cosmological effects such as redshift space distortions, these will propagate into our adapted catalogues making them a robust reflection of a realistic clustering-based redshift experiment.

From the S$^3$-SAX catalogue we can construct two samples (explained in Sections \ref{OpticalCat} and \ref{SimulatingHIIntensity} respectively) which we will refer to as 

\begin{itemize}[leftmargin=*]
\item Optical galaxy catalogue (subscripted with g)
\item \hi intensity maps (subscripted with HI).
\end{itemize}
The optical galaxy catalogue is the catalogue that we will be treating as our sample of `unknown' redshifts, and for which we will try to recover the true redshift distribution. We will only need the galaxy positions from this catalogue, and from these we can construct a number density field $n_\text{g}$ by binning each galaxy into a pixel.

The intensity maps will be thin slices in chosen intervals of redshift space and as is commonly the case with intensity maps, each slice will be a field of brightness temperature $T_\text{HI}$ where regions of higher temperature indicate a higher matter density. Figure \ref{SAXfluxHistogram} shows distributions of galaxy \hi fluxes ($I_{\rm HI}$) contained within our full S$^3$-SAX catalogue.

For maps produced using the \sax simulation we use a resolution of 2 pixels per arcminute which corresponds to 720 $\times$ 720 pixels maps for our 36deg$^2$ patch of sky. We also restrict the catalogue to redshifts of $0<z<3$ and use 30 redshift bins giving bin widths of $\Delta z = 0.1$. For the number of \sax galaxies contained within these ranges this gives an average number density of 4.6 galaxies per voxel.

%----------------------------------------------------------------------------------------
\subsection{Simulating HI Intensity Maps}\label{SimulatingHIIntensity}
%----------------------------------------------------------------------------------------

While traditional optical galaxy surveys aim to resolve their targets and build a catalogue of discrete objects above some lower flux detection limit, intensity mapping instead collects flux from all sources of emission, even the very faint ones, to build a continuous map of intensity. We therefore choose not to place any limits on which \hi emitting galaxies to include in our simulation to make this as realistic as possible. In other words, every galaxy within the S$^3$-SAX catalogue that has a non-zero amount of \hi emission, regardless of how faint, is included as a contributor to our \hi intensity map. We note however that we can only include galaxies which are above the simulation completeness limit. In the case of the S$^3$-SAX catalogue, the simulation is complete for galaxies with cold hydrogen masses (\hi+ H$\,\textsc{ii}\xspace$) above $10^8 M_{\odot}$.

We express our \hi intensity map data $T_\text{HI}$ in the form of a brightness temperature with two angular dimensions ($\theta_\text{ra}$ and $\theta_\text{dec}$, jointly represented by $\vec{\theta}$) and a radial dimension which is the redshift ($z$). The intensity map can be decomposed into three different map contributions
\begin{equation}\label{IMcomponenet}
	T_{\text{HI}}(\vec{\theta},z) = s(\vec{\theta},z) + f(\vec{\theta},z) + n(\vec{\theta},z).
\end{equation}
Here $s$ represents the true \hi signal we are aiming to detect, $f$ are the radio foregrounds and $n$ is noise associated with instrument systematics. The overall aim for successful intensity mapping is to therefore isolate $s(\vec{\theta},z)$ by subtracting or minimising the other unwanted components. We will discuss each of these components further in the following sections together with our method for simulating the ``cleaned" data, i.e. the HI maps after some foreground cleaning technique has been applied.

%----------------------------------------------------------------------------------------
\subsubsection{Signal}\label{TheSignal}
%----------------------------------------------------------------------------------------

To construct our intensity mapping signal we start with the \hi mass $M_{\text{HI}}$ of each galaxy, which is estimated in the S$^3$-SAX catalogue. Note that in any case we can use the formula outlined in \citep{BattyeBINGOSingleDish} to infer $M_{\text{HI}}$ from the raw signal $S_{\text{obs}}dv$, which is the flux integrated over a velocity width to capture the full \hi signal that is stretched in frequency due to the galaxy's rotational velocity:
\begin{equation}
    M_{\text{HI}} = \frac{2.35\times10^5M_{\odot}}{1+z}\frac{S_{\text{obs}}dv}{\text{Jy km s}^{-1}}\left(\frac{d_\text{L}(z)}{\text{Mpc}}^2\right) \, .
\end{equation}
We then place our galaxies into a data cube with coordinates ($\theta_\text{ra}, \theta_\text{dec}, z$) by binning each galaxy's \hi mass into its relevant pixel so we end up with a gridded \hi mass map $M_{\text{HI}}(\vec{\theta},z_c)$.

\begin{figure*}
\centering
\subfloat[No FG Contamination]
	{\includegraphics[height=6.15cm]{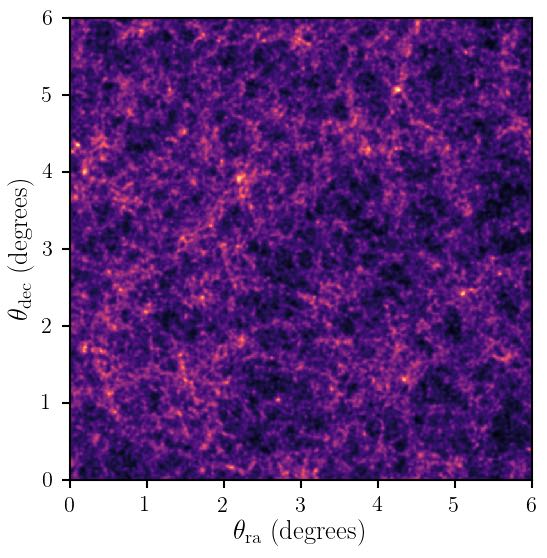}}
\subfloat[Simulated FG Clean ($\xi=0.1$)]
	{\includegraphics[height=6.15cm]{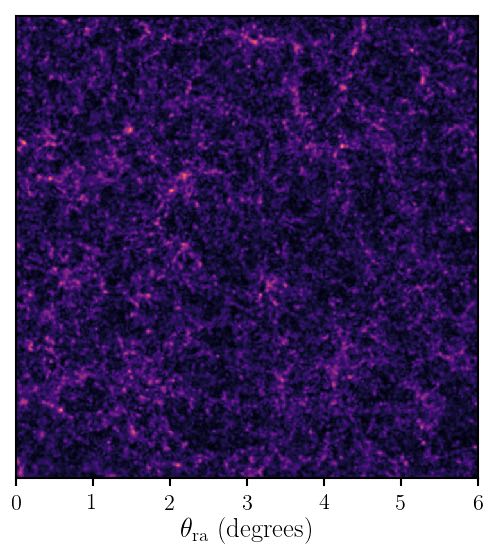}} 
\subfloat[Residuals from (a) and (b)]
	{\includegraphics[height=6.425cm]{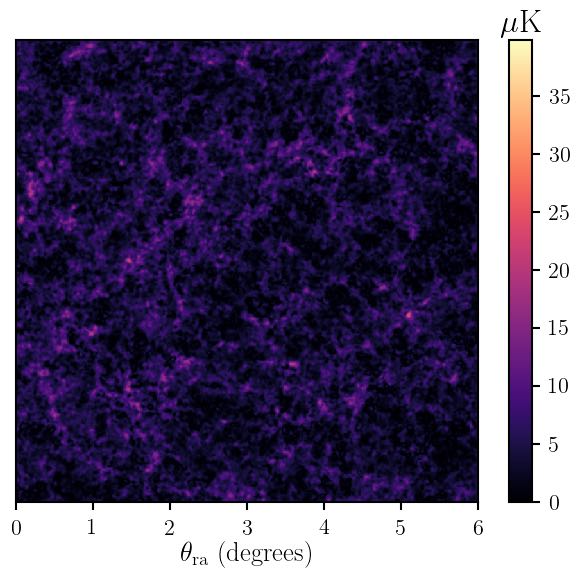}}
\caption{Example of a \hi intensity map from our simulation using S$^3$-SAX catalogue galaxies. This particular example is a slice taken at $1.3<z<1.4$ with $\theta_\text{beam}=4'$. (a) shows the raw signal with no foreground contamination, (b) shows the same signal but with some large radial modes removed from the data to simulate some of the effects of a foreground clean as explained in Section \ref{FGintro}. Differences can be seen by eye between these two but we also include the residual map (c) to clarify the impact.}
\label{IMexamples}
\end{figure*}

We can then convert this into an intensity field for a frequency width of $\delta \nu$ subtending a solid angle $\delta \Omega$ (which is effectively our pixel size)
\begin{equation}
    I_{\text{HI}}(\vec{\theta},z) = \frac{3h_\text{P}A_{12}}{16\pi m_\text{h}}\frac{1}{\left[(1+z)\chi(z)\right]^2}\frac{M_{\text{HI}}(\vec{\theta},z)}{\delta \nu \delta \Omega}\nu_{21}
\end{equation}
where $h_\text{P}$ is the Planck constant, $A_{12}$ the Einstein coefficient which quantifies the rate of spontaneous photon emission by the hydrogen atom, $m_\text{h}$ is the mass of the hydrogen atom, $\nu_{21}$ the rest frequency of the \tocm emission and $\chi(z)$ is the comoving distance out to redshift $z$ (we will assume a flat universe).

As already mentioned, it is conventional in radio astronomy, in particular intensity mapping, to use brightness temperature which can be defined as the flux density per unit solid angle of a source measured in units of equivalent blackbody temperature. Hence, our intensity $I_{\text{HI}}(\vec{\theta},z)$ can be written in terms of a black-body temperature in the Rayleigh-Jeans approximation $T=Ic^2/(2k_\text{B}\nu^2)$ where $k_\text{B}$ is the Boltzmann constant. Using this we can estimate the brightness temperature at redshift $z$
\begin{equation}
    T_{\text{HI,sig}}(\vec{\theta},z) = \frac{3h_\text{P}c^2A_{12}}{32\pi m_\text{h}k_\text{B}\nu_{21}}\frac{1}{\left[(1+z)\chi(z)\right]^2}\frac{M_{\text{HI}}(\vec{\theta},z)}{\delta \nu \delta \Omega}.
\end{equation}
Note we have used the notation $T_{\text{HI,sig}}$ to distinguish this raw signal from the true data $T_{\text{HI}}$ outlined in $\eqref{IMcomponenet}$, which includes the foreground and noise components. Lastly, to model the low angular resolution of an intensity map, we convolve $T_{\text{HI,sig}}$ with a telescope beam in Fourier space making use of the convolution theorem. Our telescope beam is modelled as a symmetric, two-dimensional Gaussian function with a full width half maximum of $\theta_\text{beam}$ acting only in the directions perpendicular to the line of sight (as the frequency/redshift resolution is excellent).

Our clustering-based redshift method will cross-correlate  optical galaxies with 2D angular intensity maps at various redshifts. We therefore choose to slice the intensity maps into thin tomographic redshift bins and collapse these to a 2D slice. The width of each tomographic redshift bin needs to be thin enough that we can make certain thin bin assumptions, yet wide enough that we allow for sufficient structure to obtain a strong cross-correlation signal. By thin bin assumptions we are referring to cosmological quantities such as the bias, which we assume to be constant within the width of our bin. This is discussed in more detail in Section \ref{EstimatorFormalism}. An example of a completed intensity map tomographically sliced and collapsed into a 2D angular map is shown in Figure \ref{IMexamples}(a).

%----------------------------------------------------------------------------------------
\subsubsection{Foregrounds }\label{FGintro}
%----------------------------------------------------------------------------------------

Arguably the biggest obstacle facing intensity mapping is the presence of  foregrounds which emit signals below 1400 MHz and which can be several orders of magnitude brighter than the \hi signal we are aiming to detect \citep{WolzForegrounds,MarioSantosForegrounds}. The term `foregrounds' is perhaps misleading as some of these contaminants do not necessarily lie in front of the \hi emitters. However, it is a term that is widely used in other literature so we will also adopt it here. The different types of foregrounds include 

\begin{itemize}[leftmargin=*]
\item Galactic synchrotron: Caused by high-energy cosmic ray
electrons accelerated by the Galactic magnetic field. This is the most dominant of the foregrounds and can be polarized or unpolarized.
\item Point sources: Emission from extragalactic point radio sources e.g. AGNs. These can potentially cluster in the same way as the \hi signal.
\item Galactic \& extragalactic free-free emission: caused by free
electrons accelerated by ions, which trace the warm ionised medium both within the Milky Way and in the broader cosmic field.
\end{itemize}
Modelling and addressing the foreground removal problem with dedicated simulations is a very active area of research (see, for example, \citep{WolzForegrounds, ShawForegrounds, AlonsoBlindForegroundSubtraction, WolzForegrounds2}). The conclusion of such work is that component separation methods can in principle be used to remove these foregrounds. The general idea is that the \hi signal spectra fluctuate in frequency whereas the foreground spectra are expected to be smooth with long frequency coherence thus making them distinguishable. However, foreground cleaning based on this approach is typically more efficient on small scales i.e. small radial modes. On larger scales the \hi signal is more similar to the foregrounds, so the result of these types of foreground cleaning can render larger radial modes useless. This has particular importance in the context of using \hi intensity maps for clustering redshift estimation \citep{AlonsoIMClusteringz} since information in these modes which could be utilised by the estimator is lost. 

In our work, rather than simulating full foreground maps, adding these onto our signal to contaminate it and then applying some removal technique, we will instead bypass this step and aim directly to simulate the aforementioned effects of foreground cleaning by removing large radial modes from the data. We follow \citet{AlonsoIMClusteringz} in imposing that any comoving radial wavenumber $k_{\parallel}$ below a certain scale $k^\text{FG}_{\parallel}$,  where
\begin{equation}\label{k_FG}
	k^\text{FG}_{\parallel} \approx \frac{\pi H(z)}{c(1+z)\xi}
\end{equation}
is rendered inaccessible by foreground cleaning; we therefore remove these modes. Here, $\xi$ parametrises the characteristic frequency scale over which foregrounds are separable from the signal. In other words a smaller value of $\xi$ means more large scale signal is lost, hence we need to remove a higher number of modes. To allow finer control of this foreground removal in fourier space we chose to increase the  resolution in the line of sight direction by splitting the redshift bins into 5 pixels each giving our \sax intensity map cube 150 pixels along the line of sight. We will investigate the effect of different values of $\xi$ in Section \ref{ForegroundRemoval}.

Hence, the recipe for simulating the effect of foreground cleaning can be summarised as \newline

\noindent(i)   Fourier transform the $T_\text{HI}$ data cube;\newline
\noindent(ii)  Eliminate (set to zero) all pixels where $k_{\parallel} < k^\text{FG}_{\parallel}$;\newline
\noindent(iii) Inverse Fourier transform back.\newline

\noindent The result of this process is an intensity map cube with some large radial modes lost. We can visualise the effects this has on the tomographic slices of intensity maps in Figure \ref{IMexamples}.

This method of simulating foreground removal is a crude approximation of the problem and we appreciate that our approach assumes all other modes above $k^\text{FG}_{\parallel}$ are cleaned with 100\% efficiency, which is of course an optimistic expectation. This has particular relevance for the auto-correlations since we expect foreground systematics to be a much bigger problem compared to the cross-correlation. However, the main issue when using intensity mapping for clustering-based redshift estimation is a loss of signal-to-noise on foreground dominated modes, rendering them ineffective when using them in correlation functions. With this in mind, we can test the main limitations of foreground cleaning by subtracting large radial modes from our reference sample.

%----------------------------------------------------------------------------------------
\subsubsection{Noise}
%----------------------------------------------------------------------------------------

Systematic effects and noise typically associated with radio surveys will impact intensity maps. Again, simulating those in detail would be a paper in its own right; for example \citet{1/fNoise} investigate the effects of $1/f$ noise (an instrumental effect that results in multiplicative gain fluctuations) in single-dish observations.

We can partially justify omitting  survey specific additive systematic effects since we would expect these to drop out in any cross-correlations between intensity maps and optical surveys. We can see this if we write 
the surveys' observable over-densities as a sum of the true signal and noise/additive systematics:
\begin{align}
	\delta_\text{g} = \delta_\text{g}^\text{signal} + \delta_\text{g}^\text{noise} \, , \\
%\end{equation}
%
%\begin{equation}
	\delta_\text{HI} = \delta_\text{HI}^\text{signal} + \delta_\text{HI}^\text{noise} \, .
\end{align}
Hence, when we cross-correlate we expect the noise terms to be uncorrelated (due to the different telescope and survey properties), leaving just the cosmological signal:
\begin{equation}
\begin{split}
	\langle\delta_\text{g}\delta_\text{HI} \rangle = \ & \langle \delta_\text{g}^\text{signal}\delta_\text{HI}^\text{signal}\rangle + \langle \delta_\text{g}^\text{signal}\delta_\text{HI}^\text{noise}\rangle \\
&+ \langle \delta_\text{g}^\text{noise} \delta_\text{HI}^\text{signal}\rangle + \langle \delta_\text{g}^\text{noise} \delta_\text{HI}^\text{noise}\rangle \\
= \ & \langle \delta_\text{g}^\text{signal}\delta_\text{HI}^\text{signal}\rangle.
\end{split}
\end{equation}
Here we have used the fact that the signal-noise cross terms are uncorrelated and that the noise maps from each survey will be uncorrelated too. Strictly speaking, while this argument is valid for the expected cross-correlation, it is not valid for the uncertainties on that cross-correlation. For example if we have some survey-specific large-scale noise, it will cancel out in the cross-correlation, but will still dominate the error budget on that cross-correlation on large scales. It is also worth noting that we make use of auto-correlations in our work too, and in these situations the above argument does not apply. However, for the purposes of this paper we assume all instrumental systematics are either negligible or drop out and do not cause any contamination in our results. We leave a full simulation involving noise maps, which will look into whether realistic telescope noise levels are sub-dominant, for future work.

%----------------------------------------------------------------------------------------
\subsection{Optical Galaxy Sample}\label{OpticalCat}
%----------------------------------------------------------------------------------------

It is important that the optical galaxy samples have realistic redshift distributions which tail off at higher redshifts where resolved detection becomes more difficult. We therefore choose not to use all galaxies in the simulated catalogue, but instead randomly exclude galaxies in each redshift bin until a model distribution is achieved. This also means that the optical galaxy redshift distribution will differ from the distribution of the galaxies which contribute to the \hi intensity maps, where we use all galaxies available. This makes for a more realistic test of this method too. For our optical model redshift distribution we use

\begin{equation}\label{dNdzModel}
	\frac{dN_\text{g}}{dz} = z^\beta \text{exp}(-(z\alpha/z_\text{m})^\gamma)
\end{equation}
where we use $\alpha = \sqrt{2}$, $\beta=2$ and $\gamma=1.5$ to make the distribution representative of a typical stage-IV optical large scale structure survey such as LSST or \euclid. For the mid-redshift parameter $z_\text{m}$ we use the mid-redshift for the particular simulated catalogue we are applying this to. For our \sax catalogue, this will be $z_\text{m} = 1.5$.

%----------------------------------------------------------------------------------------
\subsubsection{Simulated Optical Photometry}\label{PhotozSec}
%----------------------------------------------------------------------------------------

The procedure of estimating galaxy redshift distributions through \hi clustering will be sensitive to a number of additional effects outside of the intensity mapping foregrounds, many of which will stem from the complexities of the physics and data analysis turning the dark matter power spectrum into galaxy SEDs and eventually photometric redshift estimates. However, large community efforts are put into simulating these effects to allow investigation of non-linear effects on the power spectra (and biases) of various tracers, and the testing and validation of photometric redshift estimation codes (which typically produce highly non-Gaussian estimates with significant tails and catastrophic outliers) \citep[e.g.][]{MersonGALFORM, MacCrann:2018nej}. However, for the \hi clustering redshifts considered here it is necessary to simulate \hi emission which is correctly correlated with the optical emission measured by photometric surveys. This is particularly difficult as the galaxies making up much of the \hi signal in intensity maps are expected to be within  $\approx 10^9 \, h^{-1} M_{\odot}$ halos, \citep{Villaescusa-Navarro:2018vsg} orders of magnitude below the halo masses relevant (and hence simulated) for optical surveys, particularly in simulation boxes large enough to supply the wide and deep light-cones relevant to intensity mapping experiments. Given the potential utility of \hi clustering redshifts, and other interest in cross-correlation of stage-IV radio and optical surveys \citep[e.g.][]{Alonso:2015sfa, Harrison:2016stv, AlkistisIMoptCMBcross}, such a simulation is clearly needed, and we expect it to be pursued in further work.

For now, we take two approaches. For our principal results making use of the S3-SAX simulation, we defer this problem, estimating the full redshift distribution for the sample, rather than binning according to an estimated photometric redshift. As described in \cref{sec:a2a} we also investigate the ability to calibrate realistic simulated redshifts for the LSST telescope, but using \hi intensity maps which only contain \hi emission from the optically selected galaxies which we generate ourselves by using a HI-mass halo-mass relation.

%----------------------------------------------------------------------------------------
\section{Estimator Formalism}\label{EstimatorFormalism}
%----------------------------------------------------------------------------------------

In this section we discuss the formalism associated with our method and provide a step-by-step construction of the estimator we use to make redshift predictions for the `unknown' optical photometric sample.

Firstly, from the optical galaxy catalogue, we take the true galaxy redshifts and build a normalised redshift distribution given by
\begin{equation}
	\frac{dN_{\text{true}}}{dz}(z) = \frac{N_\text{g}(z)}{\sum\limits_i{N_\text{g}(z_i)}}\frac{1}{\Delta z}
\end{equation}
where $N_\text{g}(z)$ is the galaxy count in a given redshift bin. We normalise by dividing through by all galaxies in each $i$-bin and by the redshift bin width $\Delta z$. The aim of this work is to be able to recover this true redshift distribution. Our approach for doing this is to utilise angular correlation functions. We start by binning our \hi intensity map into thin tomographic redshift slices and take the observable \hi brightness temperature fluctuations $\delta T_\text{HI}$ for each slice defined as
\begin{equation}\label{deltaT}
	\delta T_\text{HI}(\vec{\theta},z) = T_\text{HI}(\vec{\theta},z) - \bar{T}_\text{HI}(z) \, ,
\end{equation}
where a barred quantity denotes the average value for the particular field. We also take the optical galaxy count overdensity $\delta_\text{g}$ for the full redshift range defined as
\begin{equation}
	\delta_\text{g}(\vec{\theta}) = \frac{n_\text{g}(\vec{\theta}) - \bar{n}_\text{g}}{\bar{n}_\text{g}}\, .
\end{equation}
We then calculate the angular cross-correlation between each \hi slice $\delta T_\text{HI}(\vec{\theta},z)$ and the unknown-redshift optical galaxy overdensity $\delta_\text{g}(\vec{\theta})$:
\begin{equation}\label{zerolag}
    w_\text{g,HI}(z) = \langle\delta_\text{g}(\vec{\theta})\delta T_\text{HI}(\vec{\theta},z)\rangle
\end{equation}
where the angled brackets signify an averaging over all positions in the field. This approach is therefore only focusing on the zero-lag of the angular correlation function, as we are only averaging over pixels in each map which share the same position $\vec{\theta}$. Previous clustering redshift works using resolved galaxy positions for both samples tend to extend beyond the zero-lag and attempt to gain more signal from the full-correlation function at extending separations. They then weight their correlation function such that it delivers the best signal-to-noise. For example, \citet{DES1clusteringz} and \citet{DES2clusteringz} average their correlation function $w$ over a separation range such that
\begin{equation}
	\bar{w}(z) = \int_{R_\text{min}}^{R_\text{max}}W(R)w(R,z)dR
\end{equation}
where $R$ is the separation distance between galaxies being correlated and $W(R)\propto R^{-1}$ is a weighting function, whose integral is normalised to unity and constructed to give higher weight to smaller scales; this maximises the signal-to-noise of the correlation function. They choose to use integration limits of 500 kpc and 1500 kpc and discuss how including larger scales tends to give a poorer signal-to-noise while smaller scales are more likely to suffer from non-linear
bias.   
 
Since we are using low resolution maps and correlating pixels rather than resolved galaxies, our choice is somewhat simplified. The low resolutions we use, which are constrained by the intensity mapping instrument's capabilities (the beam size), mean that often one or two pixels are representative of the preferred separations probed by the resolved optical galaxy clustering redshift methods. Also, given that the weight function prioritises smaller scales, the full-correlation function method will be very similar to using the zero-lag at the low resolutions we work with. We experimented with this using a maximum separation of $R_\text{max} = 1500$ kpc which, for the resolutions used on our \sax catalogue, corresponds to 1 pixel of separation for $z\geq 1.95$, 2 pixels for $0.95 < z <1.95$ and only reaching 16 and 9 pixels of separation for the lowest two redshift bins. Only very small deviations from the zero-lag approach would therefore be expected  given this and we do in fact find that the results from the two approaches converge in the regime where the full correlation function is tuned to maximise the signal-to-noise ratio. 

Where we have strong correlation we infer that the particular redshift bin is well represented in the overall redshift distribution i.e. we suppose
\begin{equation}
    \frac{dN_\text{g}}{dz}(z) \propto w_\text{g,HI}(z) \, .
\end{equation}
To understand the full version of this equation and build an estimator for $dN_g/dz$ we must consider the clustering amplitudes (bias terms), the underlying dark matter density, and the relationship between them. We can begin by looking at the $\delta_\text{g}$ and $\delta T_\text{HI}$ fields separately. Firstly, under the assumption of linear and deterministic biasing (expected to be accurate on large scales), we have
\begin{equation}
    \delta_\text{g} = \int_{0}^{z_\text{max}} b_\text{g}(z)\delta(\vec{\theta},z)\frac{dN_\text{g}}{dz}(z)dz
\end{equation}
where $b_\text{g}$ is the bias for the optical galaxies, $\delta$ is the dark matter over-density field and $dN_g/dz$ represents the normalised redshift distribution. Similarly, for the \hi brightness temperature fluctuations we have
\begin{equation}\label{deltaTintegral}
    \delta T_\text{HI} = \int_{0}^{z_\text{max}}\bar{T}_\text{HI}(z)b_\text{HI}(z)\delta(\vec{\theta},z)\frac{dN_\text{HI}}{dz}(z)dz \, .
\end{equation}
We can slice our reference intensity maps into appropriately thin redshift bins,
\begin{equation}
	\frac{dN_\text{HI}}{dz}(z) = \Theta(z_1,z_2) \, ,
\end{equation}
\begin{equation}
\Theta(z_1,z_2) =
\begin{cases}
0 & z < z_1 \\
1 & z_1\leq z\leq z_2 \\
0 & z > z_2 \, ,
\end{cases}
\end{equation}
where we have used the top-hat function $\Theta$ to take a slice of our HI intensity map. We now cross-correlate $\delta_\text{g}$ and $\delta T_\text{HI}$ for the redshift range chosen by $\Theta$,
\begin{equation}\label{firstcrosscorr}
\begin{split}
\langle\delta_\text{g}\delta T_\text{HI}\rangle =
 \iint & \bar{T}_\text{HI}(z')b_\text{g}(z)b_\text{HI}(z')\langle\delta(\vec{\theta},z)\delta(\vec{\theta},z')\rangle \\
 & \frac{dN_\text{g}}{dz}(z_\text{c}) \Theta(z_1,z_2) dz dz'.
\end{split}
\end{equation}
The top-hat function $\Theta$ restricts the integral to a thin redshift range and at this point we assume that we have picked a sufficiently thin bin width such that all terms become constant over this redshift range with central redshift $z_\text{c}$, leading to
\begin{equation}
\begin{split}
    \langle\delta_\text{g}\delta T_\text{HI}\rangle = \bar{T}_\text{HI}(z_\text{c}) & b_\text{g}(z_\text{c})b_\text{HI}(z_\text{c})\langle\delta(\vec{\theta},z_\text{c})\delta(\vec{\theta},z_\text{c})\rangle \\ 
& \frac{dN_\text{g}}{dz}(z_\text{c})\Delta z.
\end{split}
\end{equation}
Here, $\Delta z$ appears from the Limber approximation where we assume zero correlation outside the redshift range, so we just integrate over the small $dz$ segments where non-zero signal exists. $\Delta z$ therefore represents the bin width. $\langle\delta_\text{g}\delta T_\text{HI}\rangle$ is our zero-lag angular cross-correlation statistic where we average over all positions in the field as expressed in equation $\eqref{zerolag}$ i.e. $w_\text{g,HI} \equiv \langle\delta_\text{g}\delta T_\text{HI}\rangle$, so writing in this form gives
\begin{equation}\label{CrossCorr}
    w_\text{g,HI}(z_c) = \bar{T}_\text{HI}(z_\text{c})b_\text{g}(z_\text{c})b_\text{HI}(z_\text{c})w_\text{DM}(z_\text{c})\frac{dN_\text{g}}{dz}(z_\text{c})\Delta z \, ,
\end{equation}
where $w_\text{DM} = \langle\delta\delta\rangle$ is the dark matter correlation function. We can make use of the auto-correlation of the intensity maps to eliminate the dark matter density auto-correlation $w_\text{DM}$ from equation $\eqref{CrossCorr}$. This auto-correlation is derived using similar steps to those above and is given by
\begin{equation}
    w_\text{HI,HI}(z_\text{c}) = \bar{T}^2_\text{HI}(z_\text{c})b^2_\text{HI}(z_\text{c})w_\text{DM}(z_c) \, .
\end{equation}
Dividing equation $\eqref{CrossCorr}$ through by 
$ w_\text{HI,HI}(z_\text{c})$ we therefore get
\begin{equation}
    \frac{w_\text{g,HI}(z_\text{c})}{w_\text{HI,HI}(z_\text{c})} = \frac{1}{\bar{T}_\text{HI}(z_\text{c})}\frac{b_\text{g}(z_\text{c})}{b_\text{HI}(z_\text{c})}\frac{dN_\text{g}}{dz}(z_\text{c})\Delta z \, .
\end{equation}
Rearranging we get our final estimator for the redshift distribution
\begin{equation}\label{FinalEstimator}
    \frac{dN_\text{g}}{dz}(z_\text{c}) = \frac{w_\text{g,HI}(z_\text{c})}{w_\text{HI,HI}(z_\text{c})}\bar{T}_\text{HI}(z_\text{c})\frac{b_\text{HI}(z_\text{c})}{b_\text{g}(z_\text{c})}\frac{1}{\Delta z}.
\end{equation}
Since $\Delta z$ is defined and $w_\text{g,HI}$, $w_\text{HI,HI}$ can be measured, we just need to know the factor $\bar{T}_\text{HI}b_\text{HI}/b_\text{g}$ to recover our redshift distribution. 

In our simulations $\bar{T}_\text{HI}$ can easily be obtained, since we know the brightness temperature $T_\text{HI}$ from each galaxy and therefore the average brightness temperature for the map. However, in reality, the actual observable is the brightness temperature fluctuation defined in equation \eqref{deltaT}. $\bar{T}_\text{HI}$ is really an unknown quantity that needs to be inferred from our measurements. It is given by \citep{BattyeBINGOSingleDish}
\begin{equation}
\bar{T}_\text{HI} = 180\Omega_{\rm HI}(z)h\frac{(1+z)^2}{H(z)/H_0} \, {\rm mK} \, ,
\end{equation} with $\Omega_{\rm HI}$ the \hi density (abundance). In principle $\Omega_{\rm HI}$ can be measured using the auto-correlation \hi power spectrum with redshift space distortions, assuming a fixed fiducial cosmology \citep{GBTHIdetection1,AlkistisIMoptCMBcross}. This then gives a measurement of $\bar{T}_\text{HI}$. In practice this will not be straightforward due to issues such as residual foreground contamination.   
In this work for simplicity we will assume $\bar{T}_\text{HI}$ is known (or can be modelled accurately) and just use the mean of our catalogue brightness temperatures. We note that $\bar{T}_\text{HI}$ is a global quantity which is defined, and can be measured, independently of a clustering redshift experiment, unlike similar normalisations for optical clustering redshift sub-samples, which will be unique to the tracer selection of each experiment. The only remaining factor to address is therefore the bias ratio, which we discuss in the following section. 

%----------------------------------------------------------------------------------------
\subsection{Bias Treatment}\label{BiasSec}
%----------------------------------------------------------------------------------------

Since using \hi as a tracer of large scale structure as a way to explore cosmology is a relatively new concept, it is still unclear how biased this tracer is. In order to obtain the relevant factor $b_\text{HI}/b_\text{g}$, we  take the simple approach of measuring the angular auto power spectra $C_\ell$ for both the optical number density field and the intensity maps. If we restrict to the large linear scales and neglect redshift space distortions, we can obtain the bias factor through
\begin{equation}\label{biasfactor}
	\frac{b_\text{HI}(z)}{b_\text{g}(z)} = \frac{1}{\bar{T}_\text{HI}}\sqrt{\frac{C_\text{HI,HI}(\ell,z)}{C_\text{g,g}(\ell,z)}} \, .
\end{equation}
It is worth pointing out that this method uses the power spectra in each redshift bin for both intensity maps and opticals. This therefore relies on the optical galaxies being binned by redshift, which is information we are assuming is poorly constrained, so the question of circularity arises. An approach that is viable is to bin the optical galaxies using the photometric redshifts, undergo our whole clustering redshift approach with this approximate bias ratio, and then refine and repeat so that self-consistency is reached.

\begin{figure}
	\includegraphics[width=\columnwidth]{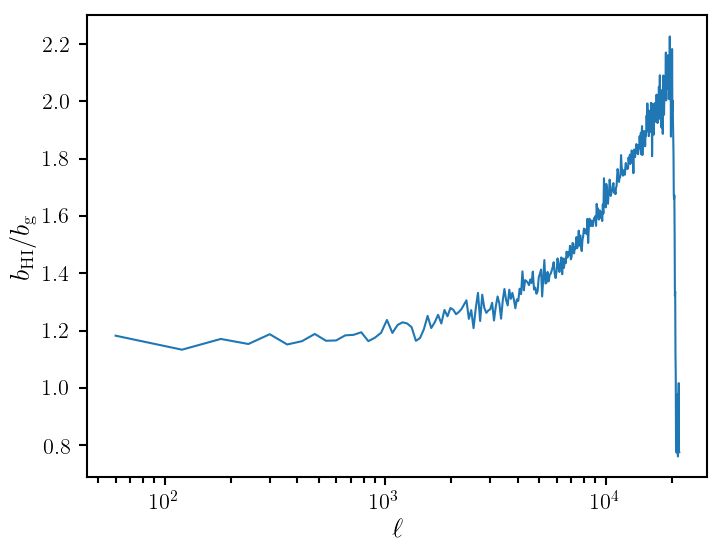}
    \caption{The bias ratio $b_{\rm HI}/b_{\rm g}$ as a function of angular scale at redshift $z=2$. This ratio is only constant on the largest scales so we therefore choose to measure this bias at scales with $\ell < 10^3$.}\label{biasratio}
\end{figure}

The exact form of the neutral hydrogen bias is an area of active research \citep{HIbiasAndTbar, VillaescusaNavarroHIbias, Castorina:2016bfm, Villaescusa-Navarro:2018vsg} and recent detections in \citet{ParkesIMxOptDetection} relied on measurements from the ALFALFA survey \citep{ALFALFA} to obtain $b_\text{HI}$ (see also the very recent work by \citet{Obuljen:2018kdy}). Furthermore, modelling bias amplitude differences between the reference and unknown samples is a problem that appears universal to clustering redshift methods. For example spectroscopic surveys cross-correlated with photometric surveys have not fully constrained these biases and offer a range of proposed solutions for addressing this in practice. In the context of this work, a further solution could be to build a model for the \hi bias through its cross-correlation with a spectroscopic or weak-lensing survey. Again, it is worth pointing out that the \hi bias may be determined independently of the clustering redshift survey, rather than in analyses where samples of optical galaxies are used, where the bias must be determined for the galaxy types making up that particular sample, which will be a function of the experiment.
For now we rely on the approach as outlined in equation \eqref{biasfactor} where we assume we can successfully obtain thin redshift slices in the optical sample and obtain perfect foreground removal (of the relevant modes) for the \hi sample. 

\begin{figure}
	\includegraphics[width=\columnwidth]{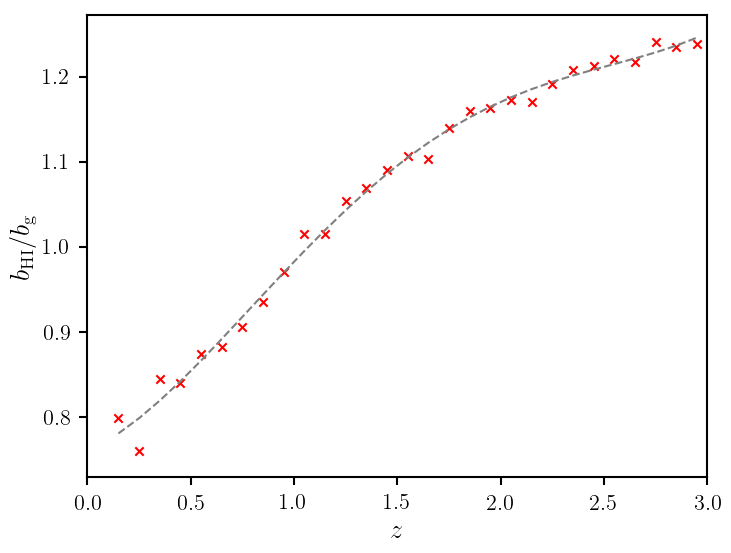}
    \caption{The bias ratio $b_{\rm HI}/b_{\rm g}$ as per equation \eqref{biasfactor} in each redshift bin with the grey dashed line showing a polynomial fit to the data points. As expected, the bias ratio that we use in our estimator evolves with redshift.}
    \label{bias_v_z}
\end{figure}

From our simulations we find that the bias factor is scale independent only at large scales, as expected. As Figure \ref{biasratio} shows for an example redshift bin, we appear to have a constant bias ratio on scales $\ell<10^3$. We find a similar relation holds in all redshift bins. We therefore chose to take the mean value for this bias ratio at the angular scales of $\ell<10^3$. Figure \ref{bias_v_z} shows how the mean value of the bias ratio used in our estimator evolves with redshift.

One final point is that in our estimator we choose to focus on the zero-lag of the correlation function, which includes small scales. However, we are only estimating the large-scale linear biases. This should not cause a problem since the small scale non-linear bias contributions are integrated out due to the low resolutions (approximately 2 pixels per arcminute) we are using. As we will see in the next section, our results show that the use of the zero-lag statistic in conjunction with large-scale linear biases appears not to  cause any issues; but consideration should be given to this point when choosing bin sizes in real survey analyses.
 
%----------------------------------------------------------------------------------------
\section{Results \& Discussion}\label{Results}
%----------------------------------------------------------------------------------------

Here we present our analysis and findings on the viability of using \hi intensity mapping for clustering-based redshift estimation. Throughout we use our estimator as laid out in Section \ref{EstimatorFormalism} and in particular equation \eqref{FinalEstimator} and proceed to investigate some of the properties that affect this method.

\begin{itemize}[leftmargin=*]
\item We begin in Section \ref{NonGaussSec} by examining the effect of \hi-bright sources on our method using basic mock-catalogues which we simulate. 
\item In Section \ref{ForegroundRemoval} we carry out the first test of our method using our adapted S$^3$-SAX catalogue (introduced in Section \ref{SimDetails}) which we construct realistic \hi intensity maps from (albeit over a small sky area) and put particular emphasis on some of the effects from foreground cleaning.
\item Section \ref{BigBeam} looks at the Gaussian beam size $\theta_\text{beam}$ and whether increasing this to realistic amounts (comparable to some single-dish experiments such as the SKA) is too damaging to our redshift predictions. This relies on extending our simulation to a larger sky area so we make use of the MICE catalogue \citep{MICE1,MICE2,MICE3,MICE4,MICE5} which has a wider light-cone than \sax. 
\item We then finish in Section \ref{sec:a2a} by looking at how this method can provide excellent information on the error associated with stage-IV photometric redshifts. For this we use simulated LSST-like photometric redshifts from \cite{AscasoA2A}. 
\end{itemize}

%---------------------------------------------------------------------------------------- 
\subsection{Bright HI-Rich Sources}\label{NonGaussSec}
%----------------------------------------------------------------------------------------

Correlation functions in conventional optical surveys consider separation between different resolved point-like positions of galaxies. For intensity mapping, where we have different intensity objects binned into pixels, care needs to be taken when computing correlation functions for fields where there is not much signal or where extremely bright sources are dominating over the rest of the signal.

\begin{figure}
  	\includegraphics[width=\columnwidth]{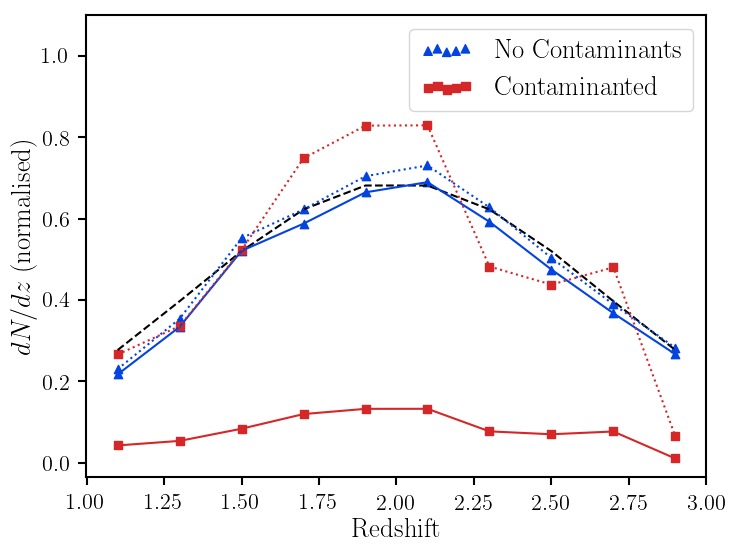}
    \caption{Mock simulation with an input redshift distribution (black dashed line) which we aim to recover. In the case where we have bright contaminating sources in our intensity maps (red square lines) our estimator struggles to recover this distribution presenting noise and scaling problems. However, results are improved when we remove these bright contaminants (blue triangle lines). The dotted lines in both cases show the results but normalised to unity to match the amplitude of the true redshift distribution which is also normalised to unity.}\label{dNdz_toy}
\end{figure}

Having a \hi-rich galaxy fall in a particular bin, whose signal vastly dominates over everything else in the field, could result in the rest of the field having essentially zero relative contribution to the signal. This can lead to the correlation function being shot-noise dominated. We want to try to avoid our fields having such extreme non-Gaussian properties, which constitute a poor representation of the underlying density field. 

We investigated the effects of this behaviour by producing mock intensity maps and then contaminated them with dominant bright sources to see how this would affect the correlation functions and impact our clustering redshift method. We did this with a simplified model where we generate galaxies with a given distribution in redshift, simulate \hi intensity maps with these galaxies, and then attempt to recover the redshift distribution with our clustering-based method. To initially ensure that no galaxy's flux was too dominant over the rest of the field we assigned all $10^7$ galaxies in our mock a uniformly random \hi flux emission between 0 and 1 (units are irrelevant for this mock example). For this simple model the input redshift distribution could be recovered since the intensity maps being produced were very uniform with Gaussian-like properties (see the blue triangle lines in Figure \ref{dNdz_toy}). However, it is possible that some galaxies will be several orders of magnitude brighter than the rest of the field as supported by the simulated fluxes from the \sax catalogue (Figure \ref{SAXfluxHistogram}).
So to investigate the effects of bright dominant sources we reassigned 1\% of the galaxies in the mock catalogue a much higher \hi flux emission, with uniformly random values between 1 and 10,000. At this point scaling problems were encountered in our mock situation along with large noise when recovering the redshift distribution, as shown in Figure \ref{dNdz_toy}. This shows that if bright sources dominate, they can contaminate the field and affect the results of the distribution recovery. The large scaling problem, shown by the red solid line in Figure \ref{dNdz_toy}, can be overcome since we are free to run a post-normalisation on the results to correct these scaling issues (shown by the dashed lines). However, the shape of the distribution still carries a large amount of noise for the contaminated case.

Therefore where possible, one should aim to avoid working with intensity maps where bright sources dominate the field and induce this extra noise in the correlation functions. An example of where this should be considered is when choosing which areas of redshift space to probe. At very low redshifts the survey volume is small, so the number of galaxies making up the intensity map is low making them more prone to bright source contamination. This in turn makes them more likely to have non-Gaussian like fields leading to a poor distribution estimation for that redshift bin. Similarly, at high redshifts the survey may only be able to detect very bright \hi sources. It is therefore imperative to choose a redshift space region, and redshift bin width, which include sufficient numbers of contributing galaxies so that one does not produce shot-noise dominated intensity maps. For this reason we exclude low redshifts ($z<0.1$) from all the catalogues we use and select a sufficient redshift bin width of either $\Delta z = 0.05$ or $0.1$ depending on redshift range of the particular catalogue. 

In reality, for intensity mapping experiments that are also performing \hi galaxy surveys like SKA, it would be possible to remove the \hi flux from a very bright source since it would likely be resolved in the \hi galaxy survey. This flux-cutting approach represents an alternative way to alleviate the problem.

%----------------------------------------------------------------------------------------
\subsection{Foreground Removal}\label{ForegroundRemoval}
%----------------------------------------------------------------------------------------

As described in Section \ref{FGintro}, a key challenge when considering using \hi intensity mapping methods for precision cosmology is foreground contamination. In this work we simulate some of the effect that foreground removal is expected to have on the recovery of the HI signal, which is to render a certain proportion of large radial modes useless. In reality, all modes would suffer some degree of foreground contamination as foreground cleaning can never separate signal and foregrounds with 100\% efficiency. But it is largely considered that the large scale modes in the line-of-sight direction are the least separable from foregrounds \citep{ShawForegrounds} and therefore these will be rendered useless.

We follow the recipe laid out in Section \ref{FGintro} and eliminate any radial wavenumber that has $k_\parallel < k^\text{FG}_\parallel$, where $k^\text{FG}_\parallel$ is defined in equation \eqref{k_FG}, to emulate the main impact of a foreground clean on our data. The $\xi$ parameter in equation \eqref{k_FG} parametrises our foreground removal whereby a lower $\xi$ equates to more radial modes being lost, signifying a harsher foreground clean.

\begin{figure}
  	\includegraphics[width=\columnwidth]{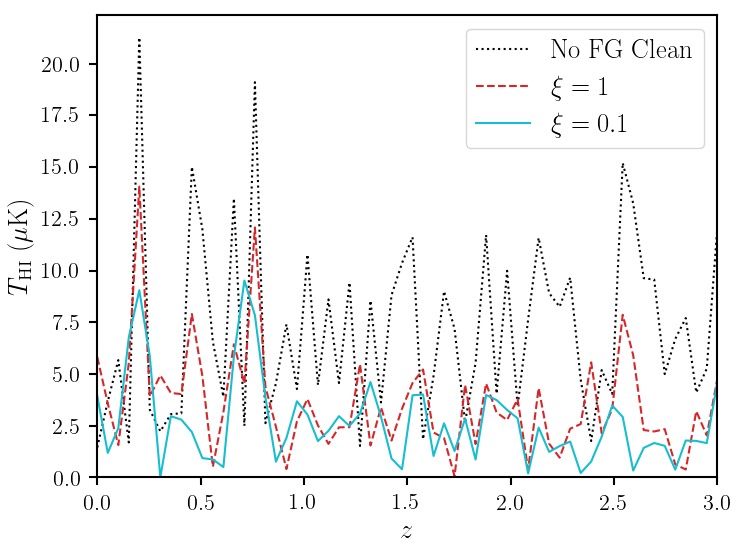}
    \caption{Demonstrates the effects of large radial mode removal (one of the effects expected from a foreground clean) and how lowering the parameter $\xi$, which translates to assuming a harsher foreground clean, gives data less representative of the original signal (the dotted black line). Done for random line of sight on our \sax catalogue.}\label{z_FGtrack}
\end{figure}

Figure \ref{z_FGtrack} shows an example of the effect that this simulated foreground removal has on a random line of sight through redshift and shows, as we expect, a suppression of the large radial modes which gets more severe for a higher $\xi$. The impact this has on the actual maps is displayed in Figure \ref{IMexamples}. The expectation is that much of the angular clustering information still remains in the smaller scale modes that are left behind, which can still be exploited for a clustering-based redshift estimation.

\begin{figure}
\centering
  \renewcommand{\thesubfigure}{b}
\subfloat[][Simulated Foreground Clean ($\xi=0.1$)]
	{\includegraphics[width=\columnwidth]{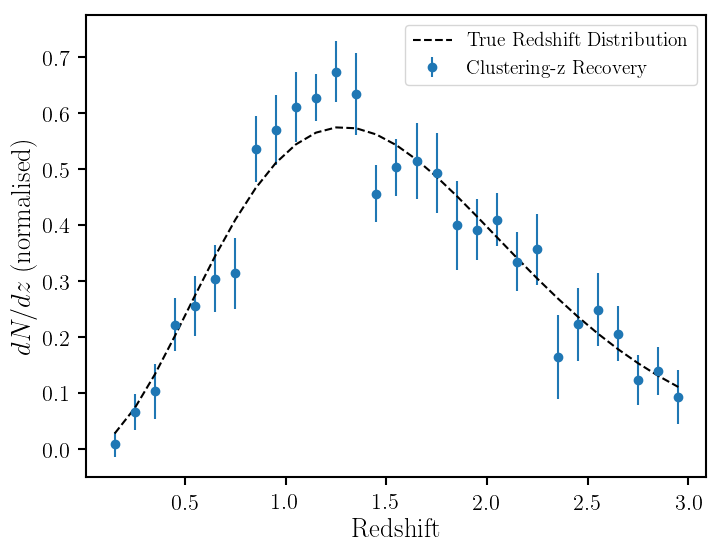}}
\caption{Results of using our \hi intensity maps to recover the redshift distribution for the `unknown' optical galaxies. The dashed lines show the true distribution which we seek to recover and the points are the estimator's prediction using a tomographic sliced intensity map at the particular redshift. Here we have used the S$^3$-SAX catalogue with $\theta_\text{beam}=4'$. (a) is the case with no foreground contamination and (b) is an example where we have applied our low-$k_\parallel$ cut with $\xi=0.1$ to simulate a foreground clean. Error bars are obtained through jackknifing over 25 samples as explained in equation \eqref{jackknife}.}
  \label{dNdz}
\end{figure}

Figure \ref{dNdz}(a) presents our first result from a redshift estimation attempt using our method on the S$^3$-SAX catalogue. For the case with no foreground contamination we find that it is still beneficial (i.e. it improves the goodness-of-fit) to nullify just one slice of pixels in $k$-space that contains the largest radial modes. This represents information at $0 < k_\parallel < 0.7 \times 10^{-3} h \text{Mpc}^{-1}$ scales and since these scales are so large, no useful information exists there to be used in the estimator's matching process. In other words these scales just contribute noise and therefore it is not surprising that their removal improves results. However, as we start to subtract more slices of pixels and eliminating information at larger values of $k_\parallel$ we get a reduction in estimator performance as desired to emulate a foreground clean. Figure \ref{dNdz}(b) shows an example with our simulated foreground clean where we have used $\xi=0.1$.

\begin{figure}
\centering
\subfloat[][No Foreground Contamination]
	{\includegraphics[width=\columnwidth]{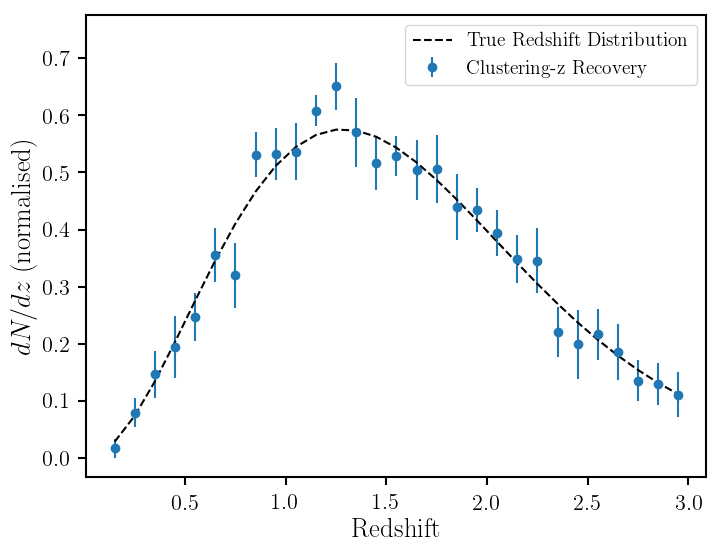}}
\end{figure}

For these plots we have used a jackknifing technique to obtain our error bars. This was done by gridding the maps into an array of $n$ smaller sub-samples, with $n=25$. We then measure our estimator, which we here denote as $\hat{x}_i$, on the map but omit the $i$-th sub-sample. We repeat the procedure, averaging over the estimators obtained from omitting sub-samples, and obtain a standard deviation via
\begin{equation}\label{jackknife}
	\sigma_\text{error} = \sqrt{\frac{n-1}{n}\sum_{i=1}^n (\hat{x}_i - \bar{\hat{x}})^2}.
\end{equation}

Figure \ref{dNdz} suggests that even with quite a harsh foreground clean, a reasonable estimation of the redshift distribution of the optical galaxies can be made. A value of $\xi \approx 0.1$ corresponds to a cut that would target more complicated foreground residuals arising from leaked polarised synchrotron. Due to Faraday rotation these would exhibit a frequency structure which is not as spectrally smooth as other foreground contaminants hence making them more likely to remain after a mode cut \citep{AlonsoIMClusteringz}.

The exact scales that are rendered inaccessible after a successful foreground clean is a subject still open for debate i.e. the most realistic value of $\xi$ is unclear. Work by \citet{ShawForegrounds} proposes a foreground cleaning method which claims to render scales with $k_{\parallel} < 0.02 \, h\text{Mpc}^{-1}$ ($\xi\approx 0.05$ at $z=1.5$) inaccessible, whereas there is more encouraging recent work by \citet{ZuoForegrounds} which suggests that foreground cleaning is possible where information from these small $k_{\parallel}$ modes may not necessarily be lost at all. They propose using an extended method, Robust Principal Component Analysis (RPCA), which utilises the sparsity of the frequency covariance for the \hi signal.

\begin{figure}
	\includegraphics[width=\columnwidth]{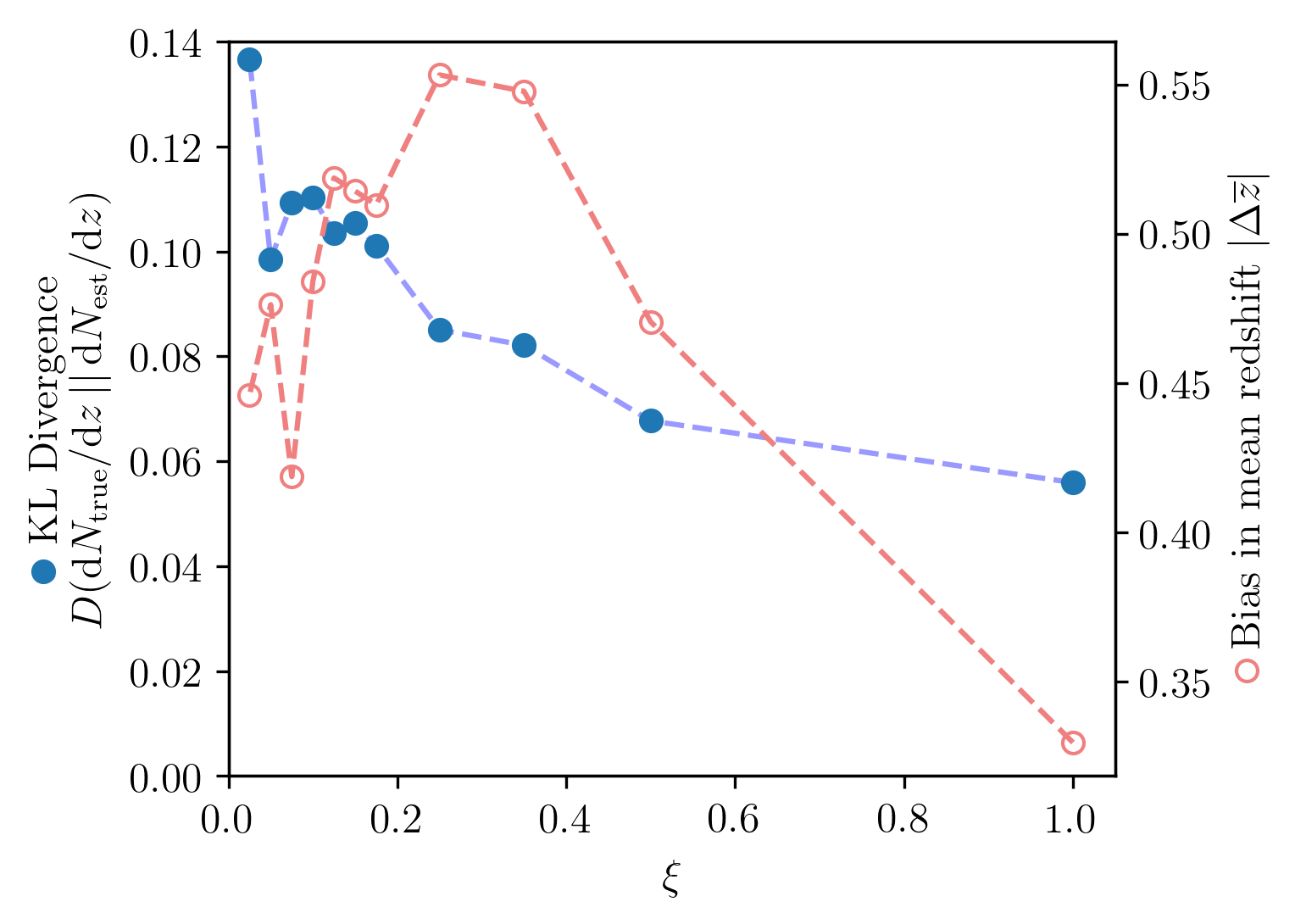}
\caption{Test of estimator performance for differing levels of foreground cleaning parametrised by $\xi$. Shown is the Kullback-Liebler divergence $D$ giving the information loss when describing the true redshift distribution with the estimated one (filled blue dots, left axis), and the bias in mean recovered for the redshift distribution (empty red dots, right axis). We see  that the ability of the clustering estimator to recover the true distribution deteriorates as we increase the amount of foreground cleaning assumed (i.e. as we decrease $\xi$).}\label{TightnessOfFit}
\end{figure}

In Figure \ref{TightnessOfFit} we examine how various values of $\xi$ affect the precision of our redshift distribution estimation, by analysing the Kullback-Liebler (KL) Divergence for different values of $\xi$ as a figure of merit. The KL divergence $D(P\,||\,Q) = \sum_i P_i \log(P_i/Q_i)$ measures the information lost when an approximating discrete distribution $Q$ is used to describe a true distribution $P$, providing a well-motivated way of estimating the goodness-of-fit across a whole distribution. Also shown is the mean recovered redshift for the distribution as a function of the same $\xi$. The plot is encouraging in showing that even when approaching conservative levels of foreground cleaning ($\xi \approx 0.1$), the degradation in performance is not significant when compared to the $\xi=1$ case.

%----------------------------------------------------------------------------------------
\subsection{Varying Beam Size}\label{BigBeam}
%----------------------------------------------------------------------------------------

\begin{figure}
	\includegraphics[width=\columnwidth]{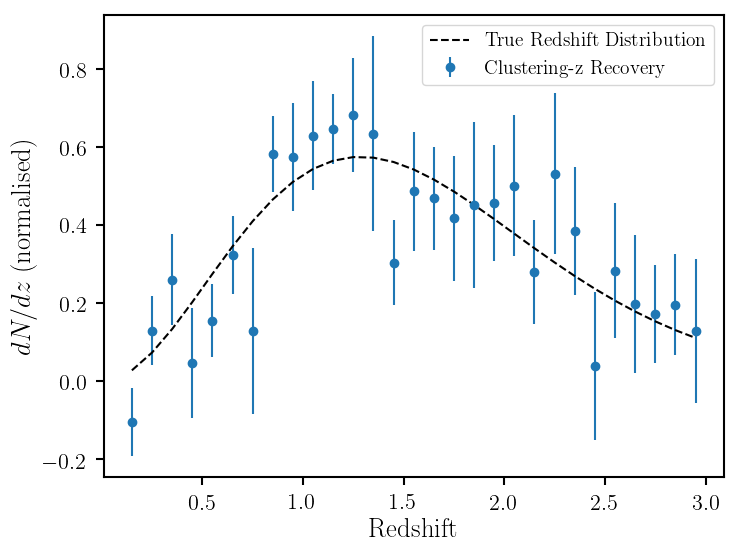}
    \caption{Increasing the beam size to $\theta_\text{beam}=16'$ for our \sax sample, which is equivalent to reducing the resolution of our experiment, causes errors to increase as predicted by equation \eqref{sigsmootherror}.}
\label{dNdz_highsmooth}
\end{figure}

Interferometric intensity mapping experiments such as CHIME (0.26$^\circ$ - 0.52$^\circ$)  \citep{CHIME} or HIRAX (0.08$^\circ$ - 0.17$^\circ$) \citep{HIRAX} have relatively good angular resolution. However, the proposed HI intensity mapping surveys using MeerKAT or SKA-MID in single-dish mode \citep{MeerKLASSMeerKAT,SKAHICosmologyPoS} are expected to have  quite large beams and therefore a low angular resolution (greater than 1.4$^\circ$). It is worth reiterating here that SKA will also operate as an interferometer, but we choose to focus on its use as a single-dish intensity mapping experiment to test the limitations of large receiver beams. In general, a single-dish intensity mapping experiment will typically have a beam size given by
\begin{equation}\label{beamequation}
	\theta_\text{beam} \approx \lambda / D_\text{dish} \, ,
\end{equation}
where $\lambda$ is the observing wavelength and $D_\text{dish}$ is the dish diameter. So for an SKA-like intensity mapping experiment in single-dish mode,  with dish diameters of $D_\text{dish}=15$m, targeting the redshifted $\lambda=21$ cm signal we would expect to have a $\theta_\text{beam}\approx 2$ deg at a median redshift of $z=1.5$. Unfortunately, for our simulations using the S$^3$-SAX catalogue we are limited to a small sky coverage of $6\times 6$ square degrees, and this limits the extent to which we can increase our beam size. Since the error on our redshift estimation $\sigma_{N(z)}$ will be inversely proportional to the square root of the number of effective pixels in our field, and the number of effective pixels will just be the area of the whole field $A$ divided by the area of our beam $\approx \theta_\text{beam}^2$, we can estimate
\begin{equation}\label{sigsmootherror}
    \sigma_{N(z)} \propto \frac{\theta_\text{beam}}{\sqrt{A}} \, .
\end{equation}
We therefore find an increase in error as we explore lower resolutions. Even with no simulated foreground clean and only increasing the beam size to $\theta_\text{beam}=16'$, we are quadrupling our error and we find a large deterioration in the precision of our prediction as shown in Figure \ref{dNdz_highsmooth}.

%----------------------------------------------------------------------------------------
\subsubsection{Testing on Larger Sky Area}\label{LargeSky}
%----------------------------------------------------------------------------------------

Because of the rapid increase in error shown in Figure \ref{dNdz_highsmooth} from increasing the beam size to $\theta_\text{beam}=16'$, we proceeded to perform a scaled up test of clustering-based redshift estimation on larger sky areas to check if we can successfully go to higher levels of $\theta_\text{beam}$. To do this we require access to a catalogue with much larger sky coverage, so we choose to use the MICE simulation \citep{MICE1,MICE2,MICE3,MICE4,MICE5}, which is a cosmological N-body dark matter only simulation resulting in a $\approx 200$ million galaxy catalogue over a $5,000$ deg$^2$ area up to a redshift $z=1.4$.

For these larger sky maps we use the HEALPix package \citep{Gorski:2004by} where the pixelation ensures that each pixel covers the same surface area as every other pixel. We handle the maps in HEALlPix RING ordering scheme with resolution $\texttt{nside} = 512$, which corresponds to $12 \times 512^2 = 3,145,728$ pixels across the full sky. Since the MICE catalogue covers angular coordinates in range $0 < \text{ra, dec} < 90$ deg, these only fill $1/8^\text{th}$ of the sky so we use 393,216 pixels for each map. 28 redshift bins are used between the redshift range of $0<z<1.4$ giving bin sizes of $\Delta z = 0.05$. For the number of MICE galaxies contained within these ranges this gives an average number density of 18.6 galaxies per voxel.

Like we did when creating our optical galaxy sample from the \sax catalogue, we use Equation \eqref{dNdzModel} as our model for an optical redshift distribution with a mid redshift of $z_\text{m}=0.7$. This creates a realistic distribution in redshift for our opticals which tails off at higher redshift and that differs from the redshift distribution of the galaxies which contribute to the \hi intensity maps.

Since this catalogue does not have apparent \hi emission-line properties for each galaxy, we must derive our own \hi masses for each galaxy. We therefore take each galaxy's halo mass as simulated by the MICE catalogue and convert this into a predicted \hi mass by following the redshift dependent prescription laid out in \citet{HImassfromHalo}
\begin{equation}\label{PadmanabhanM_HI}
	M_\text{HI} = 2N_{1}M\bigg[ \bigg(\frac{M}{M_{1}}\bigg)^{-b_{1}} + \bigg(\frac{M}{M_{1}}\bigg)^{y_{1}} \bigg]^{-1} \, ,
\end{equation}
where $M$ is the galaxy's halo mass; $M_1$, $N_1$, $b_1$ and $y_1$ are all free parameters with redshift dependence tuned to provide a best fit; we refer the reader to \citet{HImassfromHalo} for details. From this we can then follow the steps laid out in Section \ref{TheSignal} and produce mock intensity maps.

It is important to highlight that in MICE, which is primarily a simulation for optical telescopes, the halos are only resolved down to a few $10^{11}h^{-1}M_\odot$ \citep{MICE2}, and to build realistic intensity maps one would ideally want to go lower than this to ensure that \hi emission from fainter galaxies is included in the intensity maps. However, for now it is sufficient to use this catalogue to demonstrate the potential of our method; improving the mass halo resolution will primarily change the bias on our over-density field representation, which is already well sampled.

\begin{figure}	\includegraphics[width=\columnwidth]{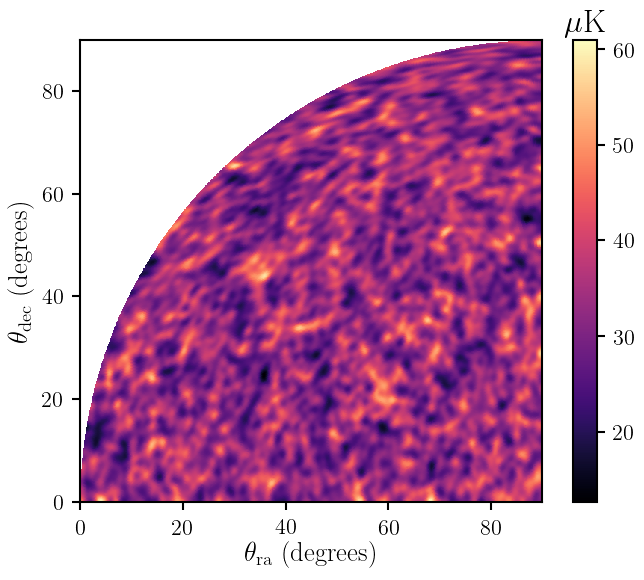}
    \caption{Large sky \hi intensity map using MICE catalogue galaxies with halo masses converted into predicted \hi masses. Since this is now a much larger patch of sky, we can no longer make the flat-sky approximation, and therefore we use a HEALPix projection for the map. This particular example is a slice taken at $0.60<z<0.65$ with $\theta_\text{beam}\approx 1.3^\circ$.}
\label{T_HIMICE}
\end{figure}

\begin{figure}
	\includegraphics[width=\columnwidth]{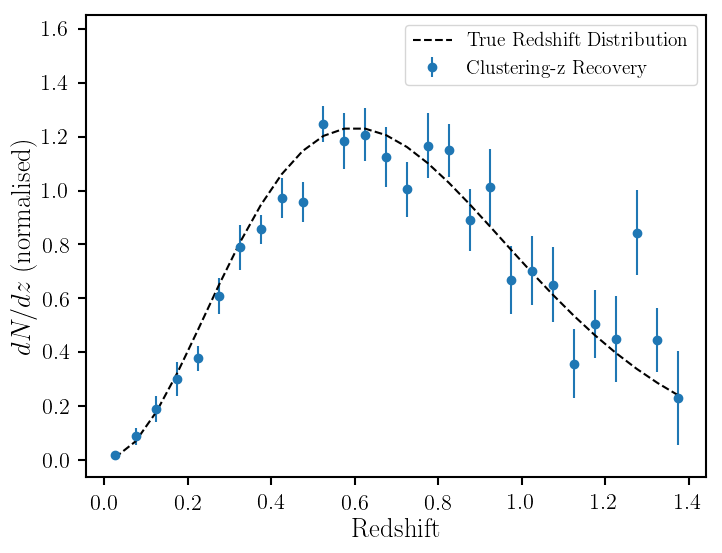}
    \caption{Results from using the large sky \hi intensity maps to recover the optical redshift distribution. Here we have used the MICE catalogue with a beam size given by Equation \eqref{beamequation} for an SKA-like single dish experiment with a dish size of $D_\text{dish}=15$m.}
\label{dNdzMICE}
\end{figure}

An example of a \hi intensity map produced from MICE is shown in Figure \ref{T_HIMICE}. Using these simulated intensity maps binned into suitable tomographic redshift slices of width $\Delta z = 0.05$, we attempt to recover the redshift distribution of an unknown optical galaxy population produced from this large sky catalogue. Figure \ref{dNdzMICE} shows the results when using an angular resolution which varies with redshift as described by \eqref{beamequation} to make the test representative of an SKA-like single-dish intensity mapping experiment beam. We also note that the increased shot noise from the higher mass cut applied to the MICE catalogue is highly sub-dominant to the beam size effect.

These results demonstrate that even with a large beam corresponding to an SKA-like single-dish HI intensity mapping experiment, an accurate redshift estimation can be made for the optical population. For cosmological HI intensity mapping surveys, telescopes may cover a sky area over $10,000 \, {\rm deg}^2$ (larger than the sky coverage from the MICE catalogue galaxies), which suggests that our results represent conservative forecasts since increased sky size should lower the errors as suggested by \eqref{sigsmootherror}. Furthermore, it is worth reiterating that intensity mapping experiments such as CHIME and HIRAX will have better angular resolution (probing angular scales as low as $0.26^\circ$ and $0.08^\circ$, respectively). 

We note that these large sky maps do not include simulated foreground cleaning due to the added complexity of not being able to use the flat-sky approximation. However, the results obtained from Section \ref{ForegroundRemoval} suggest that foreground contamination should not be a critical problem for a clustering-based redshift estimation with intensity maps.

%----------------------------------------------------------------------------------------
\subsection{Improvement on Photometric Redshift Measurements}
\label{sec:a2a}
%----------------------------------------------------------------------------------------

\begin{figure}
\centering
\includegraphics[width=\columnwidth]{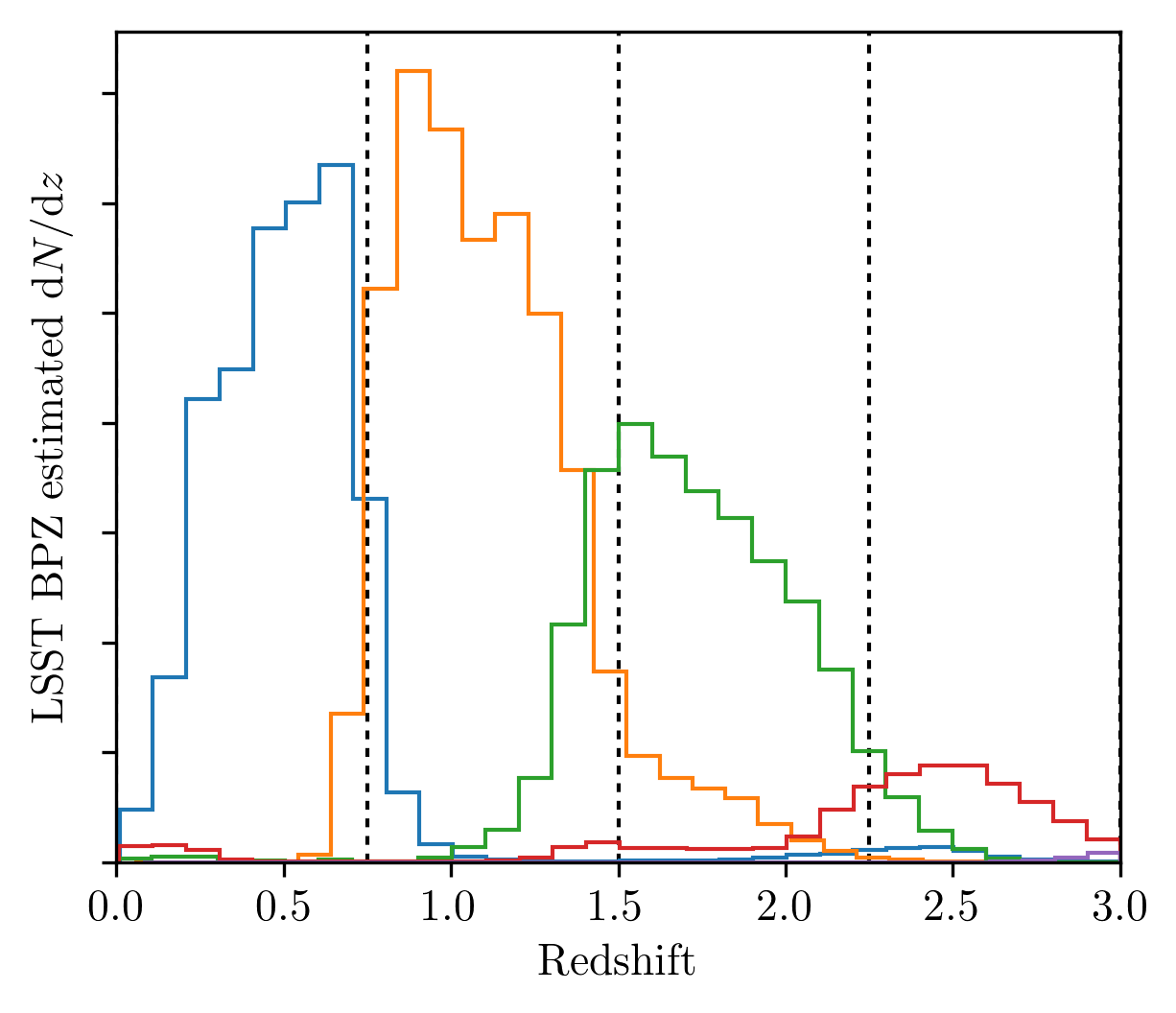}
\caption{
%The performance of LSST-like photometric redshifts simulated using the \atoa catalogue. Here galaxies with true redshifts in bins indicated by the dashed lines are separated according to their BPZ-estimated redshifts from simulated LSST photometry, with the resulting distributions shown as coloured histograms. Note the long tails and multiple modes present in all four redshift bins.
The performance of a simple redshift estimation with LSST bands from the \atoa catalogue. Here galaxies are binned (into the four bins indicated by vertical dashed lines) according to their most likely estimated redshift from running BPZ, with the histograms being of their true redshifts. This is equivalent to stacking $P(z)$ for individual galaxies in the case of Gaussian $P(z)$ with widths given by the BPZ widths.
}
  \label{A2APhotozPerformance}
\end{figure}

\begin{figure*}
\centering
	\includegraphics[width=2.1\columnwidth]{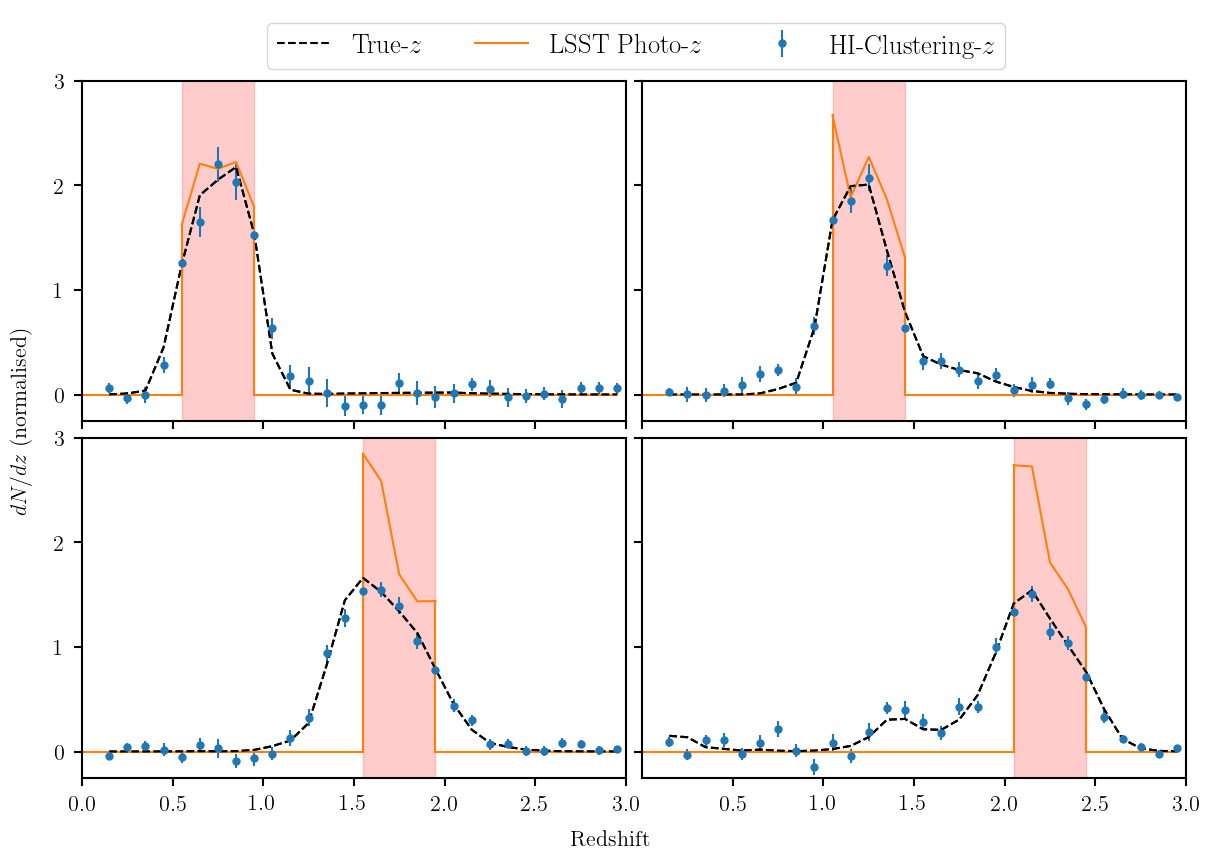}
    \caption{Complement to Figure \ref{A2APhotozPerformance} where we are now selecting galaxies based on their photometric redshift estimates. The pink shaded regions show the range in photometric redshift which galaxies are selected from. The orange line shows the distribution of these chosen galaxies according to their BPZ photometric redshift from LSST bands. The black-dashed line shows the true distribution, and the blue points show our \hi clustering redshift estimate. This was done using the \atoa catalogue adapted to include \hi emission information using equation \eqref{PadmanabhanM_HI}. Given the small sky area, intensity map resolution was set to $\theta_\text{beam}=2'$.}
\label{PhotozError}
\end{figure*}

The main aim of this work is to offer a new way of improving upon photometric redshifts, which is a major challenge for upcoming stage-IV optical telescopes like \euclid and LSST. In order to investigate this and build a pipeline we need a catalogue of galaxies for which there are robustly simulated photometric redshifts. As discussed in Section \ref{PhotozSec}, in lieu of a simulation containing all the ingredients we would like, we choose to use a simulation that has robustly simulated photometry and then add \hi emission to all galaxies using an analytical formula. Since we wish to emphasise the value of using our clustering redshift technique on future stage-IV surveys, we choose to make use of the simulated photometric redshifts from \cite{AscasoA2A} (\atoa). This includes simulated LSST and \emph{Euclid}-like photometry from the mock catalogues generated by \cite{MersonGALFORM} using the GALFORM semi-analytic code on light-cones extracted from the Millennium Simulation \citep{SpringelMillennium}. The \atoa catalogues also include photometric redshift estimates obtained using the BPZ estimation code \cite{BenitezBPZ}, which is what we use when comparing a redshift distribution obtained using photometric redshifts against our clustering-based method.

Firstly, we show the performance that we can expect from an LSST-like experiment when trying to estimate the redshift distribution using photometric redshifts. This is displayed in Figure \ref{A2APhotozPerformance}, and it shows significant deviation from the true redshift distribution. Here we bin the photometric galaxies using most-likely photometric redshift estimates obtained using the BPZ estimation code. Of course in reality LSST redshift catalogues will involve calibrations of and improvements over raw BPZ redshifts from the LSST bands, but here we simply seek to show how HI intensity mapping calibration can be one of these methods.

The \atoa catalogue we use extends to redshift $z=3$ and covers a sky area of just over 25 deg$^2$. To simulate the \hi mass for each galaxy we use equation \eqref{PadmanabhanM_HI} again as we did for the MICE catalogue in Section \ref{LargeSky}. From this we can again follow the steps laid out in Section \ref{TheSignal} and produce mock intensity maps. The \atoa simulation has a mass resolution of $1.72\E^{10} \, h^{-1}M_{\odot}$, which as discussed earlier means the simulated \hi emission will not include faint \hi emitters. This lack of completeness in our simulated intensity maps is not ideal but is likely to cause results to be worse than if we had more complete intensity maps; these would be a better representation of the underlying mass density and hence improve the precision of the correlation functions. 

With only 2,950,025 galaxies in our \atoa catalogue, an angular resolution which is identical to our SAX simulation of 2 pixels per arcminute and 30 redshift bins over a $0<z<3$ range with $\Delta = 0.1$, this gives a low number density of galaxies of 0.27 galaxies per voxel. Despite this a clustering redshift recovery is still possible.

One way of demonstrating the improvements we can make in constraining this distribution is to select a sub-population of galaxies between chosen photometric redshift limits. We can then examine the accuracy of the redshift distribution inferred from our clustering redshift method for this sub-population.

The pink shaded regions in Figure \ref{PhotozError} show various redshift intervals from which we are aiming to select galaxies. We select the galaxies using their known photometric redshifts and display their distribution with the orange line. From the photometric information alone one would conclude that a suitable population of galaxies has been selected with the desired redshift range. However, the black dashed line shows the true distribution, which extends significantly outside the claimed redshift interval. With our \hi clustering redshift method we can estimate this true distribution thus allowing the experiment to calibrate the error on the photometric selection accurately. Figure \ref{PhotozError} highlights both the need for methods that calibrate the photometric redshifts, and the potential success which our approach can have in providing this.

A further speculative approach, which is unlikely to go beyond a thought experiment level due to computational cost, would be to explore selecting galaxies from the unknown sample that maximise the correlation function signal. One could then claim that these galaxies fall within a certain redshift range based on the fact that they improve the correlation function with their inclusion. This can be put most simply by considering Figure \ref{PhotozError}. One could take the galaxies that make up the photometric redshift population as the `first-guess' for exactly which galaxies lie within the target redshift range. Then, using a sophisticated trial-and-error approach, one could remove or add galaxies that bring the true distribution (predicted by the HI-clustering redshift estimation) into agreement with the targeted redshift range. As mentioned this would be a computationally expensive process but the final result, assuming one could avoid noise contaminating the final distribution, would be a population of resolved galaxies all of which have been predicted to fall within a redshift range, which one could arguably make thin. Were this idea found to be feasible it would extend the clustering redshift method to be able to not just calibrate photometric redshift errors, but also actually improve the redshift estimates constraining them on the same scales as the bin width size ($\Delta z = 0.1$ and below).

%----------------------------------------------------------------------------------------
\section{Summary \& Conclusion}\label{Conclusion}
%----------------------------------------------------------------------------------------
\label{sec:summary}

By utilising realistic simulations of \hi emission from galaxies, we have constructed \hi intensity maps and provided evidence that they can be used to estimate the redshift distribution of a sample of optically resolved galaxies via the clustering cross-correlation method (Figure \ref{dNdz}(a)). Our estimator uses the zero-lag element of the cross-correlation function between the intensity map and optical galaxy count field, rendering it computationally inexpensive. This computational efficiency, coupled with the fact that intensity mapping will be a much faster probe compared to a spectroscopic survey, means that the method we have presented is a rapid option for constraining the redshift distribution for a large population of galaxies. Next generation surveys are promising to provide larger galaxy catalogues than ever, meaning that fast options for redshift constraints are likely to be in demand.

Given that experiments such as HIRAX, MeerKAT and the SKA have plans to operate as intensity mapping experiments in the near future, and CHIME is already taking data, a \hi clustering redshift method has particular relevance for stage-IV optical surveys such as \euclid and LSST, which will all run at similar times. While surveys such as \euclid are planning to run their own spectroscopic experiments, these are time-consuming, and LSST will be purely photometric, so in each case HI intensity mapping clustering redshifts are likely to be useful. \euclid and LSST redshift ranges are accessible to planned intensity mapping surveys such as CHIME, HIRAX, MeerKAT and the SKA and the sky overlap between many of these optical and radio surveys is excellent too. Our results from Section \ref{LargeSky} suggest that intensity mapping, even with the poor angular resolution that single-dish experiments are anticipated to have, can provide helpful redshift constraints on optical populations. It is also likely that these particular results are pessimistic since the intensity mapping experiments will most likely cover a larger sky area than that in the MICE simulation we used. Furthermore, in future the limit on halo mass resolution in simulations will decrease, emulating realistic HI intensity maps which include more faint galaxies, thus boosting the precision of the cross-correlations.

We have discussed the issue of modelling the linear bias, which is a problem that is inherent in all clustering redshift methods. This is arguably a more serious problem for in the case of \hi intensity mapping however, since the auto-correlations could potentially be further biased by contaminating foregrounds. We have made it clear that our idealistic approach of measuring the bias in our simulations would be difficult in reality; however, utilising cross-correlations with lensing data is one possible way to tackle this issue.

We also discussed some of the effects of foreground cleaning necessary for \hi intensity maps to undergo. In the context of a clustering redshift method, the largest problem this poses is that a foreground clean on intensity mapping data affects large radial modes where the foregrounds are less distinguishable from the \hi signal. We investigated this problem by removing large radial modes from our intensity maps to emulate this loss of information.

Our results, depicted in Figures \ref{dNdz}(b) and \ref{TightnessOfFit}, show that even with the loss of large portions of radial modes (low $\xi$), reasonable predictions of the redshift distribution can still be made. The fact that many large radial modes can be subtracted without too much damage to our method demonstrates that a lot of the useful matching information is in the small radial modes still exploited in the cross-correlations.

Further encouragement comes from recent work by \citet{ZuoForegrounds}, which proposes a foreground removal method that will not result in such losses of long-wavelength modes. This, together with our results, suggests that foreground contamination should not be an insurmountable problem for clustering-based redshift estimation involving \hi intensity maps.

We make this claim with a few caveats, however, as there are still aspects of the foreground problem that require further exploration. Firstly, we have only investigated the impact foreground cleaning has on large radial modes. Our method of simulating a foreground clean represents a basic approach to what is a very complex problem. It is true that foregrounds can also affect smaller scales similar to the beam size especially if considering impact from polarisation leakage. Furthermore, in the case of interferometers, additional complications need to be considered that are caused by the `foreground wedge'. This is an effect that renders an area of $k$-space, known as the `horizon-wedge', liable to foreground contamination that can be picked up from antennae with far side-lobe responses \citep{FGwedge}. In future work we will incorporate simulated foreground maps into our intensity maps and then proceed with a foreground cleaning algorithm; this is the only way to provide a fully realistic test of the effects of foreground removal.

Using simulated photometric redshifts from the \atoa catalogue we highlighted the potential improvements that could be made using clustering redshift estimation, as shown in Figure \ref{PhotozError}. This plot summarises the main point of the paper since it identifies that photometric redshifts have limitations in accuracy (especially at higher redshift) signalling the need for some accurate method of calibration, which clustering-based redshift estimation with \hi intensity mapping offers.

Producing this work has also highlighted the need for catalogue simulations capable of being used to build realistic intensity maps, which also include simulated optical photometry, and cover a large sky area. This has been discussed throughout but we reiterate that a simulation which included

\begin{itemize}[leftmargin=*]
\item simulated photometry for optically resolved galaxies so estimates using photometric redshifts can be done;
\item simulated \hi information for each galaxy for simulating realistic intensity maps;
\item low halo-mass resolution ($\approx 10^9 h^{-1}M_\odot$) so intensity maps include integrated \hi emission from faint galaxies;
\item large sky-coverage ($\approx 10,000 \, {\rm deg}^2$) to allow for testing low resolutions associated with a typical intensity mapping experiment's beam size
\end{itemize}
would be hugely beneficial not just for extending upon this work, but also for further exploration of potential synergies between optical and radio surveys.

The absence of such a simulation was significant when we extended our method to larger sky areas and quantified photometric redshift improvements. We used MICE and \atoa respectively and settled for generating our own \hi emission for each galaxy using an analytical formula (equation \eqref{PadmanabhanM_HI}). Both of these catalogues however do not have sufficient halo mass resolution for realistic \hi intensity maps. We have argued that this is only a limitation on current simulated tests and there is no reason to suppose that this will have over-inflated the effectiveness of this method. On the contrary, it is likely that obtaining lower mass-resolution, more complete \hi intensity maps would improve our results since the more realistic intensity maps would be a closer representation of the underlying mass density providing the potential for more precise correlation functions.

Given that we are expecting huge increases in galaxy number densities from upcoming galaxy surveys, the strain placed on spectroscopic follow-up is also going to increase, therefore motivating clustering-based redshift estimation methods. We believe that using \hi intensity maps within such clustering redshift methods provides an exciting possibility that warrants further investigation.

%----------------------------------------------------------------------------------------
\section*{Acknowledgements}
%----------------------------------------------------------------------------------------
It is a pleasure to thank David Alonso, Philip Bull, and Laura Wolz for very useful discussions and feedback. We also extend our gratitude to the referee whose report improved the quality of this paper. SC is supported by the University of Portsmouth. AP's work for this project was partly supported by a Dennis Sciama Fellowship at the University of Portsmouth. DB is supported by STFC consolidated grant ST/N000668/1. 

%\bibliography{Bib}
%\bibliographystyle{mn2e_plus_arxiv}

%\appendix
%\section{Validation of simulated optical photometry}
%\label{app:photo-validation}

\label{lastpage}

\end{document}